\documentclass[fleqn,usenatbib]{mnras}
\usepackage{newtxtext,newtxmath}

\usepackage[T1]{fontenc}

\DeclareRobustCommand{\VAN}[3]{#2}
\let\VANthebibliography\thebibliography
\def\thebibliography{\DeclareRobustCommand{\VAN}[3]{##3}\VANthebibliography}

\usepackage{graphicx}	
\usepackage{amsmath}	
\usepackage{multirow}
\DeclareUnicodeCharacter{2212}{-}

\title[Dark Matter and MOND successes and degeneracies]{Dark Matter and MOND successes and degeneracies in reproducing SPARC rotation curves and weak lensing data}

\author[A. Smith et al.]{Amy Smith$^{1}$\thanks{E-mail: ans1n22@soton.ac.uk},
Francesco Shankar$^{1}$\thanks{E-mail: f.shankar@soton.ac.uk},
Daeun Jeong$^{2}$,
Nicola Menci$^{3}$,
Andrea Lapi$^{4}$,
Hao Fu$^{5}$,
\newauthor
Xinyu Dai $^{6}$,
Mariangela Bernardi $^{7}$,
Arianna Di Cintio $^{8,}$$^{9}$, 
Annagrazia Puglisi$^{1}$\thanks{Anniversary Fellow},
Alexander Belyaev$^{1}$,
\newauthor
Hendrik Ulbricht$^{1}$,
Ivette Fuentes-Guridi$^{1}$
\\
$^{1}$School of Physics and Astronomy, University of Southampton, Highfield, Southampton, SO17 1BJ, UK\\
$^{2}$Astronomy Program, Department of Physics and Astronomy, Seoul National University, 1 Gwanak-ro, Gwanak-gu, Seoul 08826, Republic of Korea \\
$^{3}$ SISSA, Via Bonomea 265, 34136 Trieste, Italy \\
$^{4}$ INAF - Osservatorio Astronomico di Roma, via Frascati 33, I-00078 Monteporzio, Italy \\
$^{5}$Center for Astronomy and Astrophysics and Department of Physics, Fudan University, Shanghai 200438, China \\
$^{6}$ Homer L. Dodge Department of Physics and Astronomy, The University of Oklahoma, 440 W. Brooks St., Norman, OK 73019y \\
$^{7}$ Department of Physics and Astronomy, University of Pennsylvania, Philadelphia, PA 19104, USA \\
$^{8}$ Universidad de La Laguna, Avda. Astrofísico Fco. Sánchez s/n, E-38206 La Laguna, Tenerife, Spain \\
$^{9}$ Instituto de Astrofísica de Canarias (IAC), Calle Via Láctea s/n, E-38205 La Laguna, Tenerife, Spain
}

\date{Accepted XXX. Received YYY; in original form ZZZ}

\pubyear{2025}

\begin{document}
\label{firstpage}
\pagerange{\pageref{firstpage}--\pageref{lastpage}}
\maketitle
\begin{abstract}
The observed rotation velocities of spiral galaxies at large scales exceed those predicted by their baryonic components, motivating the need for dark matter (DM) or Modified Newtonian Dynamics (MOND). We fit the rotation curves of 147 late-type galaxies from the \texttt{SPARC} database, using a suite of models including Cold Dark Matter (CDM) models with and without baryonic feedback, Warm DM, Fuzzy DM, and MOND. We also combine \texttt{SPARC} kinematics with weak-lensing data to extend the mean rotation curves to $\sim1$ Mpc. On a galaxy-by-galaxy basis, the inner \texttt{SPARC} rotation curves marginally favour CDM models with baryonic feedback, although accounting for the uncertainties in distances and inclinations increases the degeneracy among DM-based models and MOND. All DM models predict host halo masses larger than abundance-matching expectations, with only the CDM model with baryonic feedback reducing the scatter around the mean luminosity-halo mass relation. MOND is able to reproduce the flatness of the mean galactic rotation curves on scales up to 1 Mpc, providing stellar mass-to-light ratios within observational uncertainties. In contrast, all DM models, without the contribution of mass external to the bound halo, struggle to explain the flat rotation curves. The inclusion of the current weak-lensing data tends to favour MOND over DM models, at least at the largest scales. Future observations from \texttt{BIG-SPARC}, the Square Kilometre Array, and most importantly, upcoming weak-lensing surveys such as \texttt{EUCLID} and \texttt{Rubin} will provide more stringent and independent tests of these competing frameworks.

\end{abstract}

\begin{keywords}
galaxies: haloes -- (cosmology:) dark matter-- galaxies: luminosity function, mass function
\end{keywords}



\section{Introduction}

Newtonian physics predicts that the orbital velocities at large galactocentric distances of stars in a disc galaxy should decrease proportionally to $1/\sqrt{r}$. The total velocities of disc galaxies, on the other hand, are observed to stay flat or even rise at large scales, implying the need for excess mass or a revision of Newtonian Dynamics. Dark matter (DM) was introduced to account for the "missing mass" that results in the observed flat rotation curves (RCs) seen in disc galaxies \citep[e.g.,][]{1972ApJ...176..315R,1975ApJ...201..327R,1976ComAp...6..105R,1978PhDT.......195B,1985ApJ...295..305V,1986RSPTA.320..447V,1987AJ.....93..816K}. Since then, it has been invoked to explain a wider range of astrophysical phenomena, including gravitational lensing \citep[e.g.,][]{1987A&A...184L...7S,1989ApJ...344..637G,1989ApJ...336....1L}, large-scale structure formation \citep{1978MNRAS.183..341W,2006Natur.440.1137S} and the stability of galaxy discs \citep{1973ApJ...186..467O}. The most widely accepted model is the $\mathrm{\Lambda}$ Cold Dark Matter ($\mathrm{\Lambda}$CDM) paradigm \citep{1990Natur.348..705E}, which postulates that DM consists of non-relativistic, collisionless particles that only interact gravitationally. While CDM has been successful in explaining many cosmological observations on large cosmological scales \citep{1984Natur.311..517B,1985ApJ...292..371D}, the model faces significant challenges in accounting for phenomena on smaller galactic scales, such as the core-cusp problem in spiral-galaxies \citep[e.g.,][]{2001ApJ...552L..23D,2004IAUS..220..311G,2010AdAst2010E...5D,2011AJ....142...24O}, and the satellite problem \citep{1998ARA&A..36..435M}. These limitations have prompted the development of alternative DM models, including Warm Dark Matter (WDM), Fuzzy dark matter (FDM), as well as many others \citep{2001ApJ...556...93B,PhysRevLett.85.1158}. WDM suggests that dark matter particles have a moderately higher kinetic energy compared to those of CDM, and have masses ranging between $1-10$ keV. Recent analysis of the Lyman-$\alpha$ Forest Power Spectrum have constrained the WDM particle mass to have a lower limit of $m_{\chi}\geq3.1$ keV to a $95\%$ confidence level \citep{2023PhRvD.108b3502V}. The small particle mass results in the damping of small scale structures, solving phenomena such as the missing satellite problem \citep{2001ApJ...556...93B}. Another alternative to the $\mathrm{\Lambda}$CDM paradigm is Fuzzy dark matter \citep[FDM,][]{PhysRevLett.85.1158}, which proposes that dark matter is composed of ultra-light axion-like particles with masses of $m_{\chi}\geq 10^{-22}$ eV, and a wavelength of galactic scales ($\sim1\,$kpc). The ultra-light particles show quantum behaviour at galactic scales which leads to the damping of sub-haloes \citep{2000PhRvL..85.1158H}. Another dark matter model of note is that of self-interacting dark matter (SIDM), where dark matter particles experience non-negligible elastic scattering \citep{2016MNRAS.460.1399V,2018PhR...730....1T}. These interactions redistribute energy and momentum within dark matter haloes, potentially alleviating small-scale tensions such as the core-cusp problem. In this work we will consider WDM and FDM as two typical examples of popular alternative dark matter models, which we will carefully evaluate on the same footing as CDM with and without baryonic feedback. 

Modifications to gravity such as Modified Newtonian Dynamics (MOND) have been proposed as solutions to the missing mass problem \citep{1983ApJ...270..365M}. MOND is a modification of Newtonian Dynamics at low acceleration scales, which acts to reconcile the tension between the baryonic and total circular velocities of disc galaxies in their outer regions, by essentially acting as a phantom mass distribution. MOND has been found to predict phenomena \citep[e.g.,][]{1983ApJ...270..371M,1983ApJ...270..365M,1994AnPhy.229..384M} such as the flat RCs of disc galaxies, the Baryonic Tully-Fisher Relation \citep{2000ApJ...533L..99M, 2001ApJ...563..694V}, and the Radial Acceleration Relation \citep{2016PhRvL.117t1101M, 2017ApJ...836..152L}. As for dark matter theories, MOND also presents some tensions with observations, such as the mass discrepancy in galaxy clusters \citep[e.g.,][]{2003MNRAS.342..901S,2006MNRAS.368..171Z,2017A&A...598A.127H}. In particular MOND has been tested in the case of the galaxy cluster 1E 0657-56 \citep{2006MNRAS.371..138A}, and although the relativistic MOND theory \citep[TeVeS,][]{2004PhRvD..70h3509B} can better explain the multi-centred baryonic system, it still requires excess mass that has not yet been observed, which may consist of baryonic matter \citep[e.g.,][]{2008NewAR..51..906M,2024A&A...688A..78K,2026PhRvD.113d3027Z}.

More recently, \citet{2024ApJ...969L...3M}, further extending the kpc-scale results of \citet{2021A&A...650A.113B}, analysed the stacked weak lensing data of a sample of local isolated galaxies from the KiDS survey \citep{2019A&A...625A...2K} in different bins of baryonic mass from $\log_{10}(M_{\mathrm{b}})=9\;M_{\odot}$ to $\log_{10}(M_{\mathrm{b}})=11.5\;M_{\odot}$. Via de-projection of the excess surface density of each lens, they retrieved the gravitational acceleration and thus circular velocity as a function of distance up to 1 Mpc. This scale is comparable or even surpasses the virial radii of the dark matter haloes expected to host such galaxies, favouring a more extended distribution of dark matter than predicted by numerical simulations. \citet{2024ApJ...969L...3M} suggested that MOND can naturally reproduce such flat RCs out to the largest scales, while CDM-based models would struggle in matching these data.

In this work we carefully revisit these topics focusing on the relative performance in fitting galactic rotation curves of various models, namely $\mathrm{\Lambda}$CDM with and without baryonic feedback effects, WDM, FDM, and MOND. In particular, we employ rotation curve and photometry data from the \texttt{SPARC} database \citep{Lelli_2016} to explore the performance of different models for a sample of $147$ late-type spiral galaxies. This work introduces several original elements: i) We compare on the same footing a broader, uniformly treated set of models including $\Lambda$CDM, alternative DM models, and multiple MOND formulations with different interpolation functions. ii) Instead of relying on uncertain stellar mass estimates, we use the $3.6\mathrm{\mu m}$ luminosity function instead of the stellar mass function in the abundance matching procedure, enabling a more direct and consistent comparison with SPARC data and adding an additional dimension for testing halo predictions. iii) We combine the small-scale HI/H$\alpha$ and large-scale weak-lensing \citep{2024ApJ...969L...3M} rotation curve data, providing complementary probes across a wide range of scales. Many prior studies focused either on fitting individual RCs, often optimizing model parameters on an individual basis or, alternatively, on assessing whether galaxy populations obey global scaling relations, but rarely both in a unified analysis \citep[as explored in][]{2017MNRAS.466.1648K,2018A&A...615A...3L,2019MNRAS.482.5106L,2020ApJS..247...31L,2021A&A...646L..13L,2022A&A...665A.143L}. Here we expand on \citet{2017MNRAS.466.1648K} comparing individual rotation curve parameters with global scaling relations, for a large subset of dark matter and MOND models. We then extend the analysis of galactic rotation curves out to 1 Mpc using the addition of weak lensing data from \citet{2024ApJ...969L...3M}.

Throughout this work we adopt a $\mathrm{\Lambda}$CDM cosmological framework by assuming $H_{0} = 73\:\mathrm{km s}^{-1}\:\mathrm{Mpc}^{-1}$ (as chosen in \texttt{SPARC)}, $\Omega_{\mathrm{m}}=0.3$, $\Omega_{\mathrm{b}}=0.06$, and $\Omega_{\mathrm{\Lambda}}=0.7$. 

\section{Data Selection}

This study investigates the dynamics of $147$ local late-type disc galaxies from the \texttt{SPARC} dataset \citep{Lelli_2016}, which comprises $3.6\mathrm{\mu m}$ band photometry from the Spitzer Space Telescope, along with high-precision HI and H$\mathrm{\alpha}$ rotation curve data compiled from various independent studies. The \texttt{SPARC} galaxies span a wide range in luminosity, $10^{7}-10^{12} \: L_{\odot}$, in rotational velocity, $20-300 \: \mathrm{kms^{-1}}$, and in morphological type from S0 to Irr. 
To ensure robust parameter estimation during rotation curve fitting, we selected only galaxies with the highest-quality data. Each galaxy in the \texttt{SPARC} sample is assigned a quality flag (Q), where $Q=1$ indicates high-quality HI data, $Q=2$ represents galaxies with minor asymmetries or lower-quality HI data, and $Q=3$ corresponds to galaxies with major asymmetries. $12$ galaxies with $Q=3$ were excluded, as the density profiles employed in this analysis do not account for non-circular motions, as seen in the $Q=3$ group. 
Galaxies with inclination angles $i<30$ were also excluded, as the inclination correction for observed velocities becomes unreliable at such low angles. Moreover, we excluded data points where negative gas components resulted in non-physical negative square roots during the calculation of total velocity (Equation~\ref{eq:Vtot}) in the model fitting process. The sign of the velocities indicate the direction of the gas gravitational contribution, when this is negative it indicates that particles at this point experience a stronger gravitational pull from gas at larger radii, than from gas at smaller radii. These exclusions primarily affected the innermost radial points for a small subset of galaxies. We required galaxies to have more than five total data points to ensure that each model can be constrained. After applying these selection criteria, the final sample comprises $147$ late-type disc galaxies.

To construct the extended RCs out to 1 Mpc we attach the weak lensing data from \citet{2024ApJ...969L...3M} to the mean RCs from our $147$ galaxy sample, more specifically we use the weak lensing data for all galaxy-types. Although the weak-lensing curves of late-type galaxies show on average lower velocities than early-type galaxies, we have independently checked that this does not change the results of this work significantly. We select the weak lensing data points out to 1 Mpc, as the data beyond this scale may not satisfy the isolation criterion. We also exclude the inner two data point of each weak lensing RC due to unreliable projections of mass densities. When constructing the mean RCs of \texttt{SPARC} galaxies, we only consider the mean velocities found at radial bins where three or more galaxies at measured.

\section{Method: Rotation Curve Fitting and Abundance Matching}

Below, we first describe the mass density profiles that enter each model considered in this work (Section~\ref{sec:Profiles}), the details of the statistical analysis (Section~\ref{MCMC}), and the technique of abundance matching we follow (Section~\ref{sec:HAM}).

\subsection{Halo Density profiles} \label{sec:Profiles}

Here we provide the main features of each dark matter mass density profile considered in this work. The total circular velocity of each galaxy as a function of distance from its centre, $r$, is calculated as the sum of different components

\begin{equation} \label{eq:Vtot}
    V_{\mathrm{tot}}(r) = \sqrt{\Upsilon_{*}V_{*}(r)^{2} + |V_{\mathrm{g}}(r)|V_{\mathrm{g}}(r) + 1.4\Upsilon_{*}V_{\mathrm{b}}(r)^{2}+ V_{\mathrm{DM}}(r)^{2}}
\end{equation}
where $\Upsilon_{*}$ is the stellar mass-to-light ratio, which we assume to have a Gaussian prior around $\Upsilon_{*}=0.5$ $ \mathrm{M_{\odot}/L_{\odot}} \pm 0.11$ dex as suggested by \citet{2016AJ....152..157L} and \citet{2016ApJ...816L..14L} (see Table~\ref{table:free params}) which, we checked, improves the reduced chi-square statistic compared to a flat prior. $V_{*}$, $V_{\mathrm{g}}$, and $V_{\mathrm{b}}$ are the stellar, gas, and bulge velocity components respectively, taken directly from the \texttt{SPARC} database. A bulge mass-to-light ratio of $\Upsilon_{\mathrm{b}}= 1.4\Upsilon_{*}$ is adopted here, consistent with stellar population synthesis models of old, metal-rich populations and with dynamical calibrations in the $3.6\; \mathrm{\mu m}$-band \citep{2014PASA...31...36S}.

The dark matter velocity component, $V_{\mathrm{DM}}$, is computed using the assumption of spherical symmetry and the Newtonian circular velocity equation given by

\begin{equation}\label{eq:1}
    V_{\mathrm{DM}}(r) = \sqrt{\frac{GM_{\mathrm{DM}}(r)}{r}}.
\end{equation}

As seen in equation~\ref{eq:1}, the circular velocity is a direct expression of the dark matter mass density profile. Below, we describe the dark matter density profiles that we consider in this work.

\subsubsection{Cold Dark Matter Density profiles}

One of the most popular dark matter density profiles is the cuspy Navarro-Frenk-White (NFW) profile \citep{Navarro_1996} which is a fit to the haloes found in cosmological CDM N-body simulations. The NFW-profile is characterized by two parameters: the scale density, $\rho_{\mathrm{s}}$, and the scale radius, $r_{\mathrm{s}}$

\begin{equation} \label{eq:NFW}
    \rho_{\mathrm{NFW}}(r) = \frac{\rho_{\mathrm{s}}}{\frac{r}{r_{\mathrm{s}}}\left(1+\frac{r}{r_{\mathrm{s}}}\right)^{2}},
\end{equation}
where $r_{\mathrm{s}}$ is the halo scale radius connected to the virial radius via the concentration $c_{\mathrm{vir}}$, i.e., $r_{\mathrm{s}} = R_{\mathrm{vir}}/c_{\mathrm{vir}}$. The virial radius $R_{\mathrm{vir}}$ competing to a given halo mass $M_{\mathrm{vir}}$ is computed as

\begin{equation} \label{eq:Rvir}
    R_{\mathrm{vir}}=\left(\frac{3 M_{\mathrm{vir}}}{4\pi\Delta_{\mathrm{c}}\rho_{\mathrm{c}}}\right)^\frac{1}{3}.
\end{equation}
where $\rho_{\mathrm{c}}$ and $\Delta_{\mathrm{c}}$ are the critical density and critical overdensity of the Universe. By integrating the density profile from Equation~\ref{eq:NFW} we calculated $M_{vir}$ analytically via

\begin{equation}\label{eq:Mvir_nfw}
\begin{split}
    M_{\mathrm{vir,NFW}}=&\int_{0}^{R_{\mathrm{vir}}} 4\pi r^2 \rho_{\mathrm{NFW}}(r)\,dr\\
    =&4\pi \rho_{\mathrm{s}} r_{\mathrm{s}}^{3}\left[\ln({1+c})-\frac{c}{(1+c)}\right].
\end{split}     
\end{equation}

In this work we adopted the virial definition \citep{1998ApJ...495...80B} to be consistent with the results of \citet{2017MNRAS.466.1648K} which we used as a reference throughout this work. The modelled velocities are not expected to change significantly between an overdensity, $\Delta_{c}\approx100$, and $\Delta_{c}=200$, as further discussed in Section~\ref{sec:Ext}.

\citet[][DC14 hereafter]{Cintio_2014} provided an analytic fit to the dark matter mass density profile originating in cosmological hydrodynamic simulations with baryonic feedback. The DC14 density profile accounts for baryonic feedback via the stellar-halo mass relation, and is given by

\begin{equation}
    \rho_{\mathrm{DC14}}(r) = \frac{\rho_{\mathrm{s}}}{{\left( \frac{r}{r_{\mathrm{s}}}\right)^{\gamma}} \left[1+\left(\frac{r}{r_{\mathrm{s}}}\right)^{\alpha}\right]^{(\beta-\gamma)/\alpha}}
\end{equation}
where we used the concentration definition from \citet{2017MNRAS.466.1648K}, $c_{\mathrm{sph}} = c \left[1 + \left[0.00001 \exp \left(3.4 \left(X + 4.5\right)\right)\right]\right]$. This definition of $c_{\mathrm{sph}}$ differs slightly from the one given in \citet{Cintio_2014}. We choose the one reported in \citet{2017MNRAS.466.1648K} since it results in a better fit to the simulation data from \citet{Cintio_2014}. The shape parameters, $\alpha, \beta$, and $\gamma$ are defined by

\begin{equation}
\begin{split}
    \alpha &= 2.94 - \log_{10}{\left[\left(10^{\left(X+2.33\right)}\right)^{-1.08} + \left(10^{\left(X+2.33\right)}\right)^{2.29}\right]} \\
    \beta & = 4.23 + \left(1.34X\right) + \left(0.26X^{2}\right) \\
    \gamma & = -0.06 +\log_{10}\left[{\left(10^{\left(X+2.56\right)}\right)^{-0.68} + \left(10^{\left(X+2.56\right)}\right)}\right] \\
    X & = \log_{10}\left(M_{*}/M_{\mathrm{vir}}\right).\\
\end{split}
\end{equation}

The central densities, $\rho_{s}$, here and in all other dark matter profiles are normalised to yield the mass, $M_{vir}$, within the virial radius.

The above density profiles are analytical fits to N-body simulations based on $\mathrm{\Lambda}$CDM assumptions. Other than through baryonic feedback processes, density cores can also be induced via modifications to the dark matter sector. To this purpose we chose to study two additional dark matter particle models: WDM and FDM.  

\subsubsection{Warm Dark Matter Density Profile}

The WDM halo distribution also follows the NFW density profile. However, the concentration, $c_{\mathrm{vir}}$, of WDM varies from the CDM case due to the free streaming of the dark matter particle, $m_{\chi}$. Smaller $m_{\chi}$ effectively damps small-scale density fluctuations, leading to a larger deviation from the CDM Halo Mass Function (HMF) at small scales. As $M_{\mathrm{vir}}$ decreases, the number density of WDM haloes diverges from the CDM predictions. The free streaming also results in the concentration of WDM haloes being a non-monotonic function of halo mass \citep{10.1111/j.1365-2966.2012.21252.x}, in contrast to the linear relation seen in the CDM case, as described by \citet{Dutton2014}. The CDM and WDM concentration relations used in this work are given by the following set of equations

\begin{align} \label{eq:3}
         \log_{10}\left(c_{\mathrm{vir,CDM}}\right) &= 1.025 - 0.097\log_{10}\left(\frac{M_{\mathrm{vir,CDM}}}{10^{12}}\mathrm{M_{\odot}}\mathrm{h}^{-1}\right) \\
         \label{eq:4}
         c_{\mathrm{vir,WDM}} &= c_{\mathrm{vir,CDM}}\left(1 + \gamma_{1} \left(\frac{M_{\mathrm{hm}}}{M_{\mathrm{vir,WDM}}}\right)\right) ^{-\gamma_{2}}
\end{align} 
respectively, where $\gamma_{1}$ and $\gamma_{2}$ are shape parameters taken to be $15$ and $0.3$, respectively as in \citet{10.1111/j.1365-2966.2012.21252.x}, and $M_{\mathrm{hm}}$ is the half-mode mass scale, defined as the mass at which the power spectrum of WDM is half of that of CDM. $M_{\mathrm{hm}}$ is defined as

\begin{align}
    &M_{\mathrm{hm}} \approx 2.7 \times 10^{3}M_{\mathrm{fs}} \\
    &M_{\mathrm{fs}} = \frac{4\pi}{3}\rho_{\mathrm{c}} \left(\frac{\lambda_{\mathrm{fs}}^{\mathrm{eff}}}{2}\right)^{3} \\
    &\lambda_{\mathrm{fs}}^{\mathrm{eff}} = 0.049\left[\frac{m_{\mathrm{WDM}}}{\mathrm{keV}}\right]^{-1.11}\left[\frac{\Omega_{\mathrm{WDM}}}{0.25}\right]^{0.11}\left[\frac{h}{0.7}\right]^{1.22} \mathrm{Mpc\,h}^{-1}
\end{align}
where $M_{fs}$ and $\lambda_{\mathrm{fs}}^{\mathrm{eff}}$ are the free-streaming mass scale and effective free-streaming length scale respectively. We calculated the WDM concentration, $c_{\mathrm{vir, WDM}}$ via Equation \ref{eq:4} and treated $c_{\mathrm{vir, CDM}}$ as a free parameter. 

\subsubsection{Fuzzy Dark Matter Density Profile}\label{sec:FDM}

Fuzzy dark matter (FDM) consists of ultralight axion-like particles with masses of the order of $m_{\chi} \approx 10^{-23}-10^{-22} \mathrm{eV}$ that exhibit quantum-like behaviour on galactic scales due to their de Broglie wavelengths being comparable to galactic scales. FDM haloes from N-body simulations have been fitted with a solitonic core surrounded by an NFW-envelope. The solitonic core density profile \citep{Schive2014} is given by

\begin{equation}\label{eq:sol}
    \rho_{\mathrm{sol}}(r) = \frac{1.9\left(m_{\chi}/10^{-23}\mathrm{eV}\right)^{-2}\left(r_{\mathrm{c}}/\mathrm{kpc}\right)^{-4}}{[1+9.1\times10^{-2}\left(r/r_{\mathrm{c}}\right)^{2}]^{8}} \mathrm{M_{\odot}}\mathrm{pc}^{-3}
\end{equation}
where $m_{\chi}$ is the mass of the axion-like particles, and $r_{\mathrm{c}}$ is the radius of the soliton core. In this work we assume that $m_{\chi}$ is constant across the whole galactic sample and equal to $m_{\chi}=10^{-22}$ eV. The core radius is calculated from the continuity conditions described below. The solitonic core is matched to the NFW-envelope (See equation~\ref{eq:NFW}) at the transition radius, $r_{\mathrm{t}}$, so that the density and its first derivative are continuous. The overall FDM profile and conditions are summarised as

\begin{align}
    &\rho_{\mathrm{FDM}}(r) = 
    \begin{cases}
      \rho_{\mathrm{sol}}(r) &\text{if r$ \leq r_{\mathrm{t}}$}\\
      \rho_{\mathrm{NFW}}(r) &\text{if $r > r_{\mathrm{t}}$}\\ 
    \end{cases}    \\
    &\rho_{\mathrm{sol}}(r_{\mathrm{t}}) =  \rho_{\mathrm{NFW}}(r_{\mathrm{t}}) \label{eq:cont1}\\
    &\frac{d\rho_{\mathrm{sol}}(r_{\mathrm{t}})}{dr} =  \frac{d\rho_{\mathrm{NFW}}(r_{\mathrm{t}})}{dr} .    \label{eq:cont2}
\end{align}

By using these conditions one can map the NFW free parameters, $M_{\mathrm{vir}}$, and $ c_{\mathrm{vir}}$, onto the free parameter of the solitonic core, $r_{\mathrm{c}}$, by minimising the mismatch in the continuity conditions (Equations~\ref{eq:cont1} and~\ref{eq:cont2}), where an exact solution does not always exist for a given $m_{\chi}$. This mapping reduces the number of free parameters from $5$ to $4$ in total. These free parameters are $M_{\mathrm{vir}}$, $c_{\mathrm{vir}}$, $\Upsilon_{*}$, and $r_{\mathrm{t}}$, whereas the core radius, $r_{\mathrm{c}}$, is fixed by fixing the particle mass, $m_{\chi}$.

\begin{figure}
    \centering
    \includegraphics[width=0.47\textwidth]{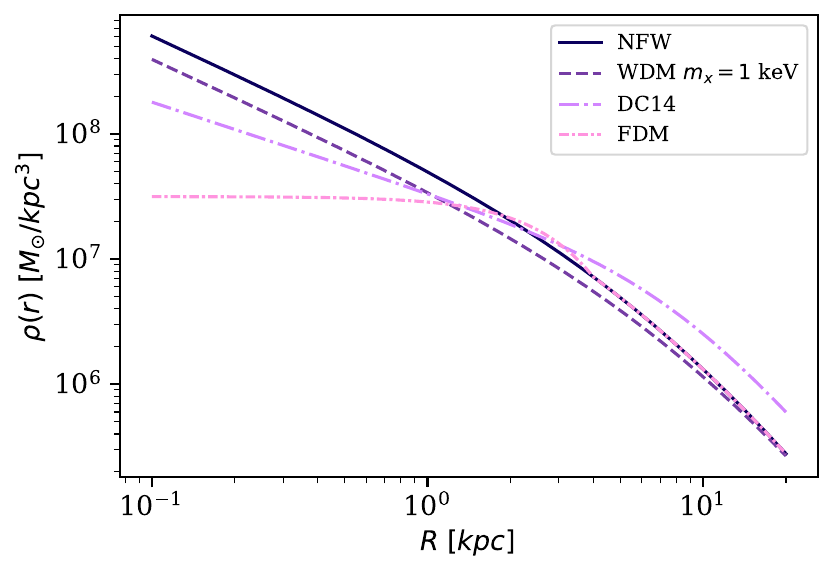}\hfill 
        \caption{The dark matter density profiles of a halo with $\log_{10}(M_{\mathrm{vir}})=11$ $\mathrm{M_{\odot}}$ and $\log_{10}(L_{[3.6]})=9.5$ $\mathrm{L_{\odot}}$. The dark blue dot-dashed, purple solid, orange densely-dashed, and yellow dashed curves plot the DC14, NFW, WDM, and FDM models respectively. The WDM model assumes a particle mass of $m_{\chi}=1$ keV. We do not show the $m_{\chi}=3$ keV case as it does not differ from the NFW curve for this particular halo.}
    \label{fig:Density Profiles} 
\end{figure}

\subsubsection{Modified Newtonian Dynamics}

Modified Newtonian Dynamics (MOND) was proposed as an alternative to dark matter \citep{1983ApJ...270..365M} to reproduce the flat RCs of galaxies that cannot be fully explained by Newtonian gravity when only visible matter is taken into account. The MOND regime becomes significant only below the critical acceleration scale, $a_{0}=1.2\times 10^{-10} \:\mathrm{m\;s}^{-2}$ \citep{1983ApJ...270..365M}. Above this scale the Newtonian regime still holds. Therefore, the effects of MOND are expected to be most apparent in the low-acceleration regions of galaxy rotation curves, specifically in the outer regions of high surface brightness (HSB) galaxies and across most radii of low surface brightness (LSB) galaxies. In this work we explore the Modified Inertia (MI) formulation of MOND, rather than the modified gravity variations such as AQUAL \citep{1984ApJ...286....7B} or QUMOND \citep{2010MNRAS.403..886M}. This choice restricts our analysis to a specific realisation of MOND in which the modification is applied to inertia rather than the gravitational field equation. Consequently, our results should be interpreted as pertaining specifically to the MI framework rather than MOND in general. The transition from the Newtonian to the MI (MOND) regime is described by an ad hoc interpolation function, $\mu(x)$

\begin{align} \label{eq:MOND}
    \mu\left(\frac{\textbf{\emph{g}}}{a_{0}}\right)\textbf{\emph{g}} = \textbf{\emph{g}}_{\mathrm{N}}
\end{align}
where $\textbf{\emph{g}}_{\mathrm{N}}$ and $\textbf{\emph{g}}$ are the Newtonian and MONDian accelerations respectively. $\mu(x)\approx x$ when $x\rightarrow 0$ and $\mu(x)\approx 1$ when $x \rightarrow \infty$. In this work we only focus on the standard (Standard IF) and simple (Simple IF) definitions of the interpolation function, given by

\begin{equation}\label{standard}
    \mu(\textbf{\emph{x}}) = \frac{\textbf{\emph{x}}}{\sqrt{1+\textbf{\emph{x}}^{2}}}
\end{equation}
and

\begin{equation}\label{simple}
    \mu(\textbf{\emph{x}}) = \frac{\textbf{\emph{x}}}{\left(1+\textbf{\emph{x}}\right)}
\end{equation}
respectively, where $\textbf{\emph{x}}=\textbf{\emph{g}}/a_{0}$. Using Equation~\ref{eq:MOND} we calculate the acceleration of each galaxy at all radial points, whilst iterating over various priors on $a_{0}$. We consider the Simple and Standard IFs of MOND under three assumptions for the acceleration scale: a fixed value of $a_{0}=1.2\times 10^{-10} \:\mathrm{m\;s}^{-2}$, a Gaussian prior centred on $a_{0}=(1.20\pm 0.02)\times 10^{-10} \:\mathrm{m\;s}^{-2}$ \citep{2016PhRvL.117t1101M}, and a flat prior over the range $a_{0}\in[0.1,4]\times 10^{-10} \:\mathrm{m\;s}^{-2}$. Although $a_{0}$ is a fundamental constant in MOND and is not expected to vary between galaxies, we deliberately allow it to vary in order to explore the impact of this assumption on parameter estimation. In particular, this serves as a controlled thought experiment to illustrate degeneracies between $a_{0}$ and other fitted parameters, rather than a test of MOND as a fundamental theory. For clarity, we retain the label "MOND" for all three cases to denote the choice of interpolation function and prior on $a_{0}$, rather than implying that the variable-$a_{0}$ case represents a physically motivated extension of the theory. In the MOND framework, the value of $a_{0}$ is determined by the normalisation of the BTFR via, $V^{4}=Ga_{0}M_{b}$ and equivalently by the acceleration scale marking the transition in the radial acceleration relation and the central density relation \citep[e.g.,][]{2000ApJ...533L..99M,2016PhRvL.117t1101M,2022NatAs...6...35L}.

From Equations~\ref{standard} and \ref{simple} we then calculate the MONDian component of the total velocity via

\begin{equation}
    V_{\mathrm{M}}(r) = \sqrt{gr}.
\end{equation}

$V_{\mathrm{M}}(r)$ is the circular velocity profile predicted by the modified inertia theory. This model has only two fitting parameters, $a_{0}$ and $\Upsilon_{*}$; or one if fixing the value of $a_{0}$. Many other works assume $a_{0}=1.2\times 10^{-10} \:\mathrm{m\;s}^{-2}$ as given in \citet{1983ApJ...270..365M} and found by \citet{2011A&A...527A..76G} using the THINGS Survey. On the other hand, other groups have found lower values, such as $a_{0} = 0.74 \times 10^{-10} \:\mathrm{m\;s}^{-2}$ from \citet{2021RAA....21..271W}, similarly to the value found from dwarf and low surface brightness (LSB) galaxies in \citet{2010ApJ...718..380S}. \citet{2018NatAs...2..668R} claimed that present disc rotation curves are inconsistent with a universal acceleration scale, which would exclude the theory of MOND altogether. However, as argued in \citet{2021A&A...646L..13L} the unavoidable degeneracies from treating $a_{0}$ as a free parameter, will artificially allow $a_{0}$ to vary from galaxy-to-galaxy. Crucially, the spread in the best-fit $a_{0}$ estimates does not reflect real physical variation, but rather the fact that $a_{0}$ is strongly degenerate with other nuisance parameters (such as $\Upsilon_{*}$, distances, and inclination corrections).

It is also useful to consider the outcomes of the External Field Effect (EFE) on the shapes of RCs. The EFE arises due to MOND being a non-linear theory of gravity (unlike Newtonian gravity), resulting in the internal accelerations of any system being susceptible to modifications via the influence of external gravitational fields. For example, a galaxy's external gravitational field can affect the accelerations of the stars within it, even if this external field is uniform and constant. This can lead to modified equations of motion that differ from the isolated MOND predictions mentioned above. These modified equations are given by \citet{2016MNRAS.458.4172H} as

\begin{equation}\label{eq:EFE}
    V_{\mathrm{StMOND}}^{2}(r) = V_{\mathrm{N}}^{2}(r)\sqrt{\frac{1-\left(\frac{a_{\mathrm{e}}}{g_{\mathrm{N}}}\right)^{2}+\sqrt{\left(1-\left(\frac{a_{\mathrm{e}}}{g_{\mathrm{N}}}\right)^{2}\right)^{2}+\frac{4(a_{0}^{2}+a_{\mathrm{e}}^{2})}{g_{\mathrm{N}}^{2}}}}{2}}
\end{equation}
for the case of the Standard IF. Here $V_{\mathrm{N}}$ is the Newtonian velocity component given by Equation~\ref{eq:Vtot} when $V_\mathrm{DM}=0$, and $a_{\mathrm{e}}$ is the external gravitational acceleration entering the MOND formalism and is treated as a free parameter in the rotation-curve fitting. It represents the MOND external field acting on the galaxy and should not be interpreted as the Newtonian external acceleration inferred from the surrounding baryonic mass distribution, since these quantities are generally not equivalent owing to the nonlinearity of MOND. We also model the Simple IF when accounting for the EFE \citep{Chae_2020}, with the total modelled velocity given by

\begin{equation}\label{eq:SimpleEFE}
    V_{\mathrm{SiMOND}}^{2}(r) = V_{\mathrm{N}}^{2}(r)\left[\frac{1}{2}-\frac{A_{\mathrm{e}}}{z}+\sqrt{\left(\frac{1}{2}-\frac{A_{\mathrm{e}}}{z}\right)^{2}+\frac{B_{\mathrm{e}}}{z}}\,\right]
\end{equation}
where $z=g_{\mathrm{N}}/a_{0}$, $e=a_{\mathrm{e}}/a_{0}$, $A_{\mathrm{e}}=e(1+e/2)/(1+e)$, and $B_{\mathrm{e}}=(1+e)$. 

The implementation of the EFE follows the one-dimensional approximation presented by \citet{2016MNRAS.458.4172H} and subsequently employed by \citet{2020ApJ...904...51C} and \citet{2022ApJ...941...55C}. The internal and external accelerations are implicitly assumed to be aligned in the same direction. As a result, $a_{\mathrm{e}}$ should be interpreted as a lower limit, although it is still expected to provide an order of magnitude estimate of the EFE's strength. In this work we compare the rotation curve fits of the Standard and Simple IFs, with and without consideration of the EFE.

\subsubsection{Distance and Inclination}

As in \citet{2021A&A...646L..13L} and \citet{2024A&A...689L...3L}, for some of our reference models we will allow the distance, $D$, to the galaxy and its inclination, $i$, to vary from their measured values in the \texttt{SPARC} database to explore the impact that observational uncertainties could have on the quality of the fits of the models. We assume a log-normal prior centred on the \texttt{SPARC} fiducial values of $D$ and $i$ with standard deviations given by their corresponding errors. The radius, $R$, and velocity of each baryonic component, $V_{x}$, are transformed via the following relations

\begin{align}\label{eq:distance}
    R' &= R\frac{D'}{D}; \quad  V_{x}' = V_{x} \sqrt{\frac{D'}{D}}.
\end{align}

The total observed rotational velocity, $V_{o}$, is modified by the inclination as shown below

\begin{align}\label{eq:inclination}
    V_{o}' &= V_{o}\frac{\sin{i}}{\sin{i'}}; \quad \delta V_{o}' = \delta V_{o}\frac{\sin{i}}{\sin{i'}}
\end{align}
where the primed variables are the modified quantities, and the non-primed variables are the fiducial \texttt{SPARC} values.

\subsection{MCMC and Parameter estimation} \label{MCMC}\label{param}

To calculate the optimal parameters for each model, we used a Monte-Carlo-Markov-Chain method, specifically we used the ensemble sampler from the Python package, \texttt{emcee} \citep{Foreman_Mackey_2013}. For each model we used $7,000$ chains which are initialised with $100$ walkers, with the stretch scale parameter set to $a=3$ as in \citet{2017MNRAS.466.1648K}, and a burn-in period of $2000$ chains. 

\begin{table} 
    \centering
    \begin{tabular}{ |p{1.6cm}||p{3.4cm}||p{2cm}|}
     \hline
     Parameter& Prior & Units\\
     \hline
     $\log_{10}(M_{\mathrm{vir}})$   & $[8,14]$    & $\mathrm{M_{\odot}}$ \\
      
     $c_{\mathrm{vir}}$&   $[1,100]$   & dimensionless\\
    
     $\Upsilon_{*}$ &$\mu=0.5$, $\sigma=0.11$ dex  & $\mathrm{M_{\odot}}/\mathrm{L_{\odot}}$\\

     $r_{\mathrm{t}}$ & $[0.01,100]$ & kpc \\

     $a_{0}$ & $[0.1,4]$ & $10^{-10}\mathrm{m\;s}^{-2}$ \\

     $a_{\mathrm{e}}$ & $[0,4]$ & $10^{-10}\mathrm{m\;s}^{-2}$ \\

     $D$ & $\mu=D_{\mathrm{SPARC}}$, $\sigma = \sigma_{D,\mathrm{SPARC}}$ & Mpc\\

     $i$ & $\mu=i_{\mathrm{SPARC}}$, $\sigma = \sigma_{i,\mathrm{SPARC}}$ & degrees\\
     
    \hline
    \end{tabular}
    \caption{Table of the parameters defining each density profile and their associated priors for MCMC parameter estimation. Here $\mu$, and $\sigma$ are the mean, and standard deviation of the log-normal distributions respectively.}
    \label{table:free params}
\end{table}

To estimate initial values of the model parameters for each MCMC run, we used non-linear least squares estimation (LSE) using the \verb'scipy.optimize.curve_fit' package and priors shown in Table \ref{table:free params}, in order to minimise the burn-in time. We minimised the residuals between the observed total velocity and the predicted total velocity of each model. For each MCMC run we used the priors as in Table \ref{table:free params}. We confirmed that using a log-normal prior on $\Upsilon_{*}$ decreased the reduced-$\chi^{2}$ compared to using a flat prior on $\Upsilon_{*}$, hence why we chose to use the log-normal prior in this work. There has been much discussion about $\Upsilon_{*}$ and whether it is a constant for all galaxies and across the radial distribution of each galaxy. This will be discussed further in Section \ref{ML}. 

We used the following log-likelihood function in the MCMC approach

\begin{equation} \label{eq:log-like}
    \ln(\mathcal{L}) = - \frac{1}{2}\sum_{i=1}^{N} { \left(\frac{(V_{i, \mathrm{obs}} - V_{i, \mathrm{pred}})^2}{\sigma_i^2} + {\ln(2\pi\sigma_{i}^{2})}\right)}
\end{equation}
where $\sigma_{i}$ is the uncertainty in the observed velocity.
From the MCMC chains we extract uncertainties of each parameters as the $16\%$, $50\%$, and $84\%$ percentiles of the distribution.

In the analysis of the extended rotation curves (see Section~\ref{sec:Ext}) we adopt the covariance matrix provided with the weak lensing data to fully account for correlated uncertainties between radial bins. This covariance is used to construct a multivariate Gaussian likelihood, ensuring that both statistical uncertainties and their correlations are consistently propagated into the model comparison. The corresponding log-likelihood function is given by

\begin{equation} \label{eq:Cov}
\begin{aligned}
\ln (\mathcal{L})
=& -\frac{1}{2}
\bigg[
(\mathbf{V}_{\mathrm{obs}} - \mathbf{V}_{\mathrm{pred}})^\mathrm{T}
\mathbf{C}^{-1}
(\mathbf{V}_{\mathrm{obs}} - \mathbf{V}_{\mathrm{pred}})\\
& + \ln \det \mathbf{C}
+ N \ln(2\pi)
\bigg]
\end{aligned}
\end{equation}
where $\mathbf{C}$ is the covariance matrix. The diagonal components in the inner regions are given by the \texttt{SPARC} HI variances taken from the error on the mean, while the outer regions are described by the statistical variances and their associated covariances from the weak lensing mean rotation curves. Together, these terms define the full covariance structure used in the likelihood evaluation.

\subsection{Halo Abundance Matching}\label{sec:HAM}

Below we will compare the best-fit halo masses retrieved from the \texttt{SPARC} RCs, with the "typical" halo masses that would be assigned to galaxies of a given stellar mass or luminosity via abundance matching. The Halo Abundance Matching (HAM) technique derives the relation between a galaxy and halo property by assuming that the relation is monotonic and requires that the cumulative abundance of each property can be equated. This technique has been used previously to derive relations between many luminous and dark matter properties \citep[e.g.,][]{2004MNRAS.353..189V,2004ApJ...609...35K,2006ApJ...647..201C,2006ApJ...643...14S,2010ApJ...710..903M,2011ApJ...742...16T,2013ApJ...764L..31K,2015ApJ...810...74A,2018AstL...44....8K,2019MNRAS.482.4364C,2022MNRAS.516.3206F}. We apply the abundance matching integral relation presented in \citet{2015ApJ...810...74A} (their Eq. 37) with a scatter of 0.1 dex (our results below do not depend on the exact chosen value of the scatter).
In this work, instead of comparing with the $M_*-M_h$ relation as in \citet{2017MNRAS.466.1648K}, we directly use the luminosity in the $3.6\mathrm{\mu m}$-band for the abundance matching to avoid any uncertain stellar mass correction between \texttt{SPARC} and the input galaxy stellar mass function. We use as a reference in the HAM the $3.6\mathrm{\mu m}$-band galaxy luminosity function (LF) of \citet{2009ApJ...697..506D} and the halo mass function (HMF) from \citet{2008ApJ...688..709T}. Following \citet{2013ApJ...770...57B} and \citet{2024MNRAS.532..177F} we also correct the HMF for the abundances of surviving sub-haloes before matching it to the LF via a proportionality factor, $C(a)$, 

\begin{equation}\label{subhalo}
    \phi_{\mathrm{s}}(M_{\mathrm{H}}) \approx \phi_{\mathrm{c}}(M_{\mathrm{H}}) C(a)\log_{10}\left(\frac{M_{\mathrm{H}}^{\mathrm{cut}}(a)}{M_{\mathrm{H}}}\right)\\
\end{equation}
with 
\begin{equation}\label{correction}    
    \log_{10}C(a) = -2.69 + 11.68a - 28.88a^{2} + 29.33a^{3} - 10.56a^{4}\\
\end{equation}
and
\begin{equation}\label{cut} 
    \log_{10}(M_{\mathrm{H}}^{\mathrm{cut}}) = 11.34 + 8.34a - 0.36a^{2} - 5.08a^{3} + 0.75a^{4}
\end{equation}
where $a=1/(1+z)$ is the scale factor, and $M_{H}^{cut}$ is the exponential cut-off halo mass, introduced to ensure the analytic HMF matches the Bolshoi N-body simulation at low masses.

For alternative dark matter models, we also apply a modification to the HMF using the general equation of \citet{2022Univ....8..476L},

\begin{equation}\label{general hmf}
\begin{split}
    &\frac{\mathrm{d}N_{X}}{\mathrm{d}M_{\mathrm{H}}dV} = \frac{\mathrm{d}N_{\mathrm{CDM}}}{\mathrm{d}M_{\mathrm{H}}\mathrm{d}V}\left[1+\alpha\left
    (\frac{M_{\mathrm{H}}^{\mathrm{s}}}{M_{\mathrm{H}}}\right)^{\beta}\right]^{-\gamma}
\end{split}
\end{equation}
where the cut-off mass $M_{\mathrm{H}}^{\mathrm{s}}$, and shape parameters, $\alpha$, $\beta$, $\gamma$ for each dark matter particle model are given in table \ref{table:HMF parameters}. The cut-off mass, and shape parameters in Table~\ref{table:HMF parameters} were derived using various numerical simulations \citep[][and references therein]{2022Univ....8..476L}. The cut-off halo mass has a different physical origin for the different dark matter scenarios. In the WDM framework, the characteristic suppression mass, $M_{\mathrm{H}}^{\mathrm{s}}$, is derived from the free-streaming of dark matter particles, which erases small-scale density fluctuations and thereby suppresses the formation of low-mass haloes \citep{2012MNRAS.424..684S}. In contrast, for FDM the suppression essentially arises from quantum pressure associated with the uncertainty principle, which counteracts gravitational collapse on scales below the Jeans length \citep{2016ApJ...818...89S}. We use the CDM, WDM, and FDM HMFs to estimate separate LHMRs for each dark matter scenario.

\begin{table}
    \centering
    \begin{tabular}
   {|c c c c c c|} 
         \hline
             \textbf{Model} & $M_{H}^{s} [M_{\odot}]$ & \textbf{$\alpha$} & \textbf{$\beta$} & \textbf{$\gamma$} & \textbf{Ref} \\
         \hline \\
         
         WDM & $1.9\times10^{10} (m_{\chi}/\mathrm{keV})^{−3.33}$ & 1.0 & 1.0 & 1.16 & 1\\
         
         FDM & $1.6\times10^{10} (m_{\chi}/10^{−22}\mathrm{eV})^{−1.33}$  &  1.0 & 1.1 & 2.2 & 2 \\
         
         \hline
    \end{tabular}
    \caption{The cut-off virial halo mass, $M_{\mathrm{H}}^{\mathrm{s}}$, and slope parameters ($\alpha$, $\beta$, $\gamma$) of the modified HMFs for the WDM, and FDM models. References: (1) \citet{2012MNRAS.424..684S}, (2) \citet{2016ApJ...818...89S}.}
    \label{table:HMF parameters}
\end{table}

\section{Results}

In this section we present the results of the dark matter and MONDian model fitting to the RCs of our sample of $147$ \texttt{SPARC} individual late-type galaxies (Section~\ref{RCs}). We then compare the retrieved best-fit halo masses with those expected from the halo abundance matching (HAM) technique (Section~\ref{LHMR}), and discuss the performance of the different DM models in reproducing the luminosity-halo mass relation (LHMR). Finally, we will compare our models with the mean extended RCs obtained from the combination of SPARC and weak lensing data to probe DM and MOND models in the deep regime at very low accelerations (Section~\ref{sec:MeanRC}). 

We analyse the residuals between the models and the data via the reduced chi-square statistic, $\chi^{2}_{\mathrm{\nu}}$, to compare the fitting abilities of each model. The $\chi^{2}_{\mathrm{\nu}}$ statistic of each galaxy in our sample is estimated as

\begin{equation}
    \chi^{2}_{\nu} = \frac{1}{\nu}\sum^{N}_{i=1}\left(\frac{V_{\mathrm{obs}}(r_{i}) - V_{\mathrm{pred}}(r_{i})}{\sigma_{i}}\right)^{2}
\end{equation}
where $\nu$ is the degrees of freedom given by the difference between the number of radial points per galaxy, $N$, and the number of parameters, $p$, in a given model ($\nu=N-p$). The sum is calculated over all measured radial points. 

\begin{figure}
    \includegraphics[width=0.47\textwidth]{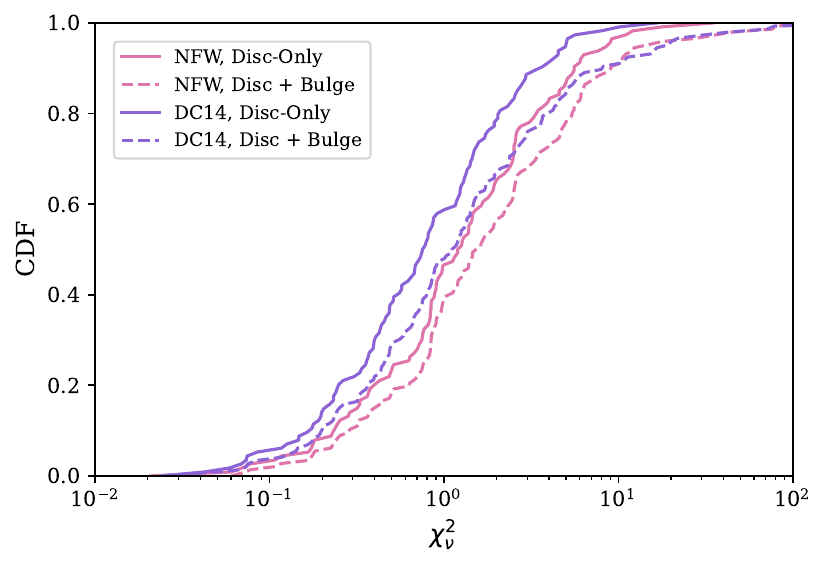}\\
    \includegraphics[width=0.46\textwidth]{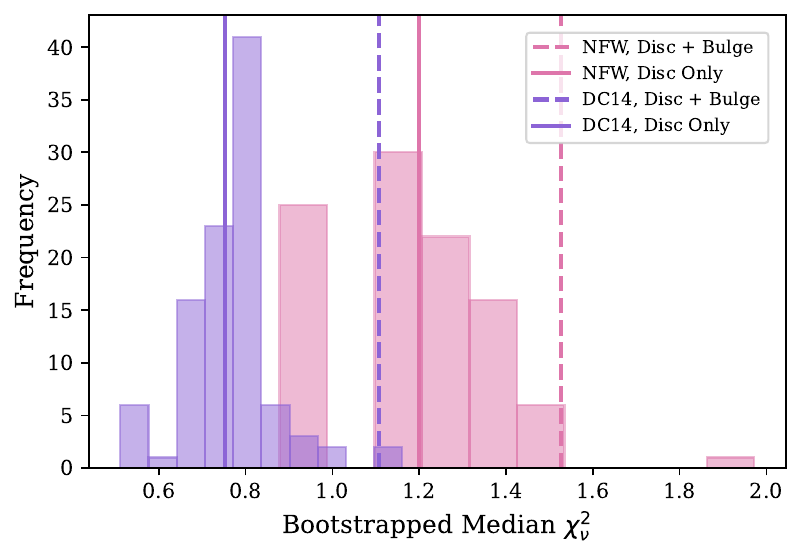}
        \caption{\textit{Top Panel: }The cumulative distribution function (CDF) of the reduced chi-square statistic, $\chi^{2}_{\mathrm{\nu}}$, for the Disc + Bulge, and Disc-Only subsamples - from rotation curve fits assuming an NFW (pink curves) and a DC14 (purple curves) density profile. The solid and dashed lines represent the $\chi^{2}_{\nu}$-CDF of the Disc-Only and the Disc + Bulge subsamples respectively. \textit{Bottom Panel: }A histogram of the median $\chi^{2}_{\mathrm{\nu}}$ frequency of $100$ randomly sampled groups of galaxies in the Disc-Only sample for the NFW (pink histogram) and DC14 (purple histogram) models, compared to the median $\chi^{2}_{\mathrm{\nu}}$ values of these subsamples (solid vertical lines). The vertical dashed lines represent the median $\chi^{2}_{\mathrm{\nu}}$ of the whole Disc + Bulge subsamples.}
    \label{fig:Chi_Bulge} 
\end{figure}

Figure~\ref{fig:Chi_Bulge} (top panel) reports our results for our two main reference models, the classical NFW profile, and the feedback-induced DC14 profile (pink and purple lines, respectively) when considering all the $147$ \texttt{SPARC} galaxies in the fitting, or the Disc-Only subsample of $115$ SPARC galaxies (dashed and solid lines, respectively). We find that including bulged galaxies has a net effect of increasing the reduced $\chi^{2}_{\mathrm{\nu}}$, anyhow, the relative performance of the two models remains roughly unchanged (see bottom panel). Since the bulge-dominated galaxies represent a relatively small population of our total \texttt{SPARC} sample, we also tested via bootstrapping whether the small size of the Bulge sample had any effect on its median $\chi^{2}_{\mathrm{\nu}}$. We took $100$ random galaxy subsamples of the size of the Bulge-Only subsample, from the Disc-Only sample and calculated their median $\chi^{2}_{\mathrm{\nu}}$ with that of the Disc + Bulge sample. The result of this exercise is included in the bottom panel of Figure~\ref{fig:Chi_Bulge}. All $100$ subsamples of the Disc-Only group resulted in median $\chi^{2}_{\mathrm{\nu}}$ values below that of the Disc + Bulge sample. Such test reveals that the morphology of the galaxies can have a non-negligible impact on the median $\chi^{2}_{\mathrm{\nu}}$ and that bulges in particular could bias the fitting towards higher $\chi^{2}_{\mathrm{\nu}}$ values. Nevertheless, as this behaviour is expected to occur systematically in all models at a similar level, in this work we choose to include both disc-only galaxies and those with bulges to ensure we cover a comprehensive and representative sample of galaxy morphologies.

\subsection{Individual Rotation Curves} \label{RCs}

\begin{figure*} 
  \centering
  \includegraphics[width=.5\linewidth]{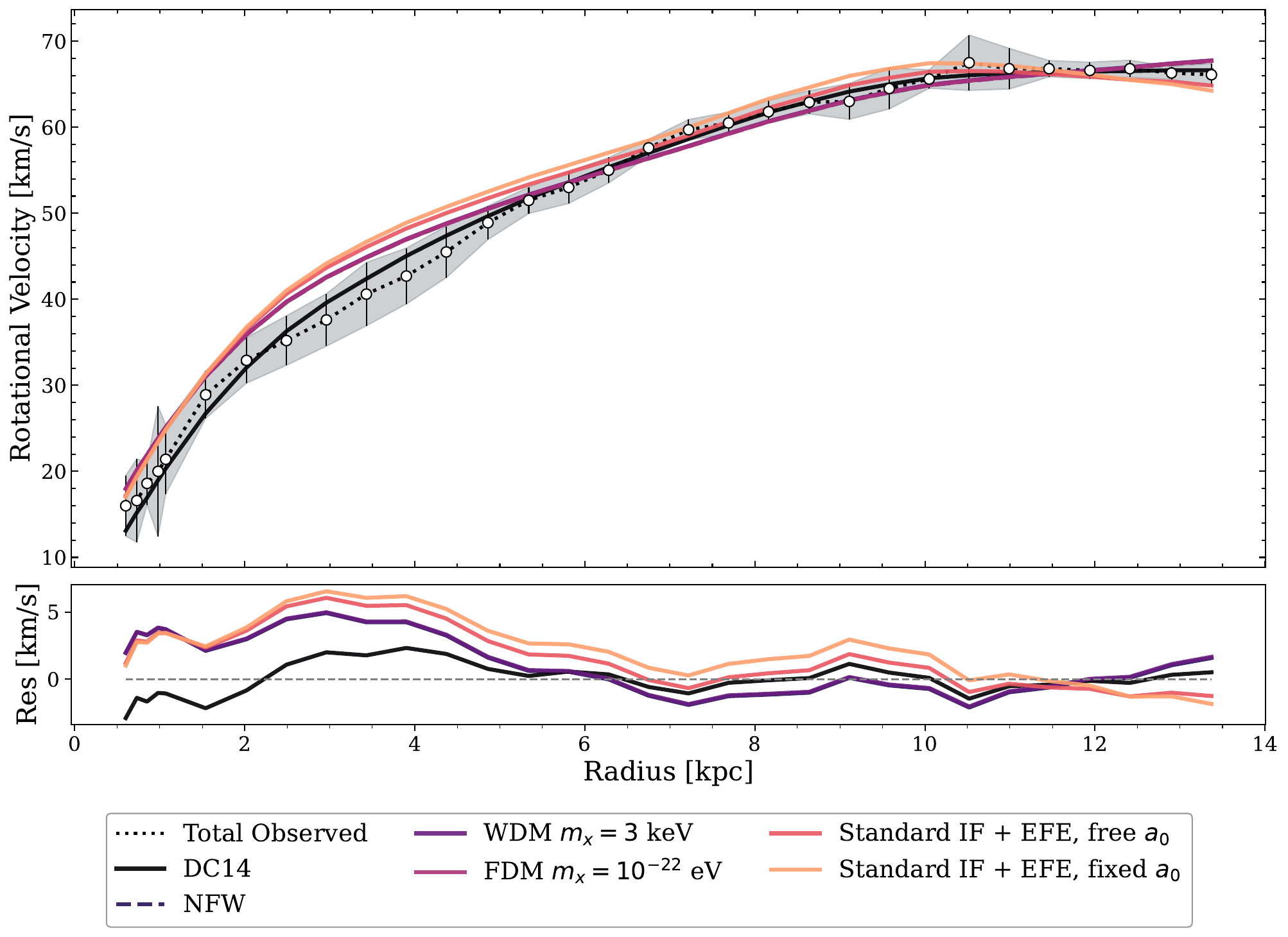}\hfill
  \includegraphics[width=.5\linewidth]{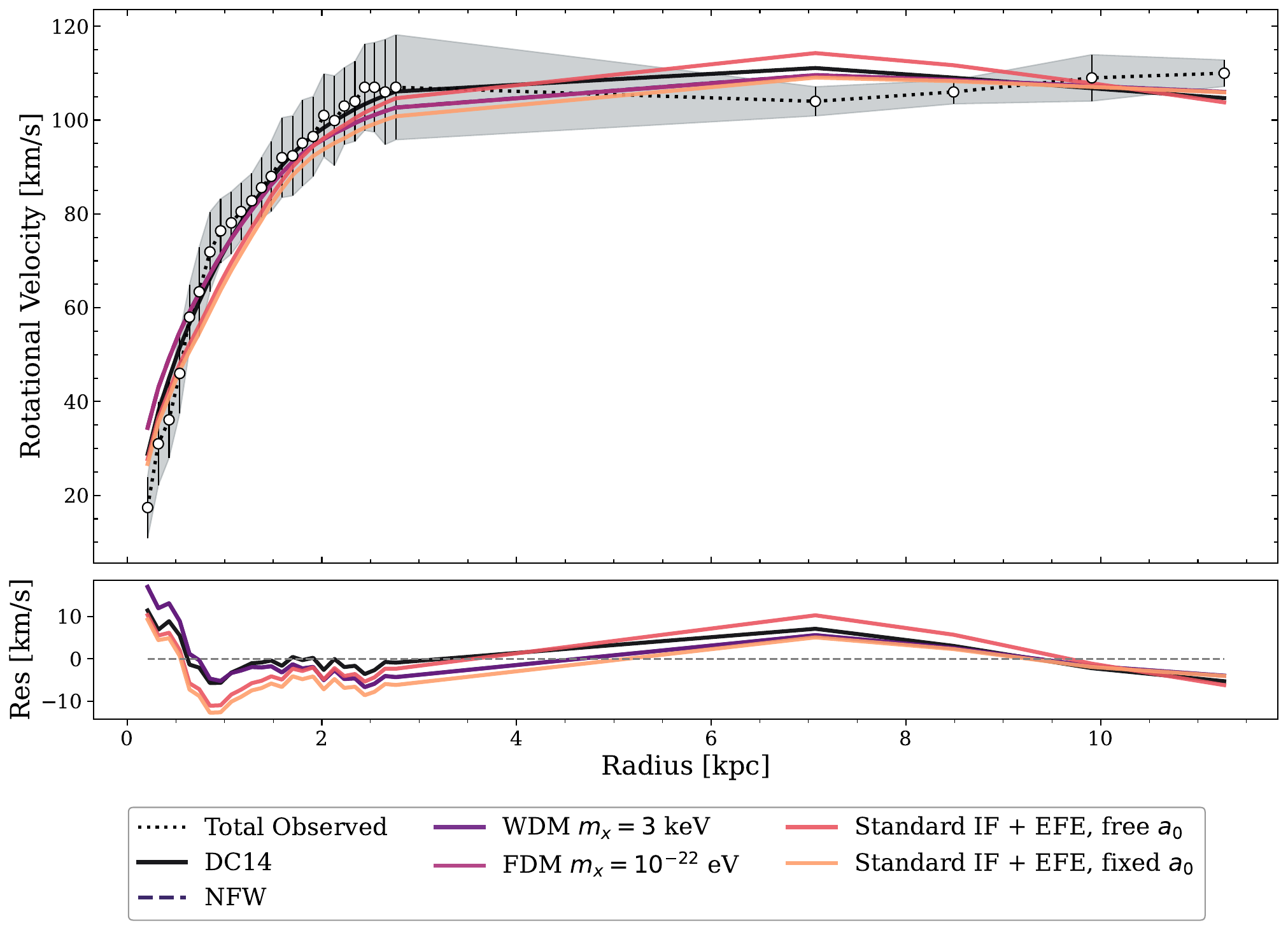}
  \caption{\textit{Left Panel:} The modelled rotation curve fits of each model (coloured curves) to the LSB galaxy,  DDO161, observed total rotation curve. \textit{Right Panel:} The modelled rotation curve fits of each model (coloured curves) to the HSB galaxy, NGC0024, observed total rotation curve. The black data points and error bars show the observed total velocities at various radii and their corresponding $1\sigma$ errors respectively. The bottom panels of each plot shows the residuals between each model prediction and the observed total velocity. }
  \label{fig:RC}
\end{figure*}

For each model, we fit the individual RCs of our $147$ galaxy sample from the \texttt{SPARC} database using the MCMC method detailed in Section~\ref{MCMC}. In Figure~\ref{fig:RC} we show two examples of the performance of different models in reproducing the RCs of two galaxies, DDO161, and NGC0024. We note that all models are able to broadly capture the whole extent of the data, showing some level of degeneracy, though the effect of the dark matter component tends to be slightly more visible in the case of lower surface brightness galaxies such as DDO161, as expected. We see that the feedback model, DC14, is able to reproduce the cored-profile of DDO161, whereas the other models slightly overestimate the velocities in this region.

It is in fact mostly the inner and outer regions where the residuals between the models and the data become more significant. A successful model should be able to capture the diversity of cusp and cored profiles in the innermost region of the galaxies, and the diversity of rising, flat, and declining velocities in the outer regions, simultaneously.
Renzo's Rule \citep{2004IAUS..220..233S} highlights the strong correlation between the shape of the rotation curve of a galaxy and the distribution of its luminous matter. Features observed in the distribution of the baryonic rotation curve are closely mirrored by features in that galaxy’s total observed rotation curve, as encapsulated by Renzo’s rule. This should be distinguished from the Universal Rotation Curve \citep{1991ApJ...368...60P,2026arXiv260427320B}, which describes average rotation curve behaviour. If there is a feature in the stellar curve not mirrored in the total observed curve it may indicate that there is a dominance of the host DM halo which smooths out the total rotation curve. The opposite also applies: if the shape of the stellar component perfectly aligns, once scaled by the $\Upsilon_{*}$, with the one measured for the total rotation curve, it may indicate a minimal contribution of DM in the inner regions at the galactic disc scale.
\begin{figure*}
    \centering
    \includegraphics[width=1\textwidth]{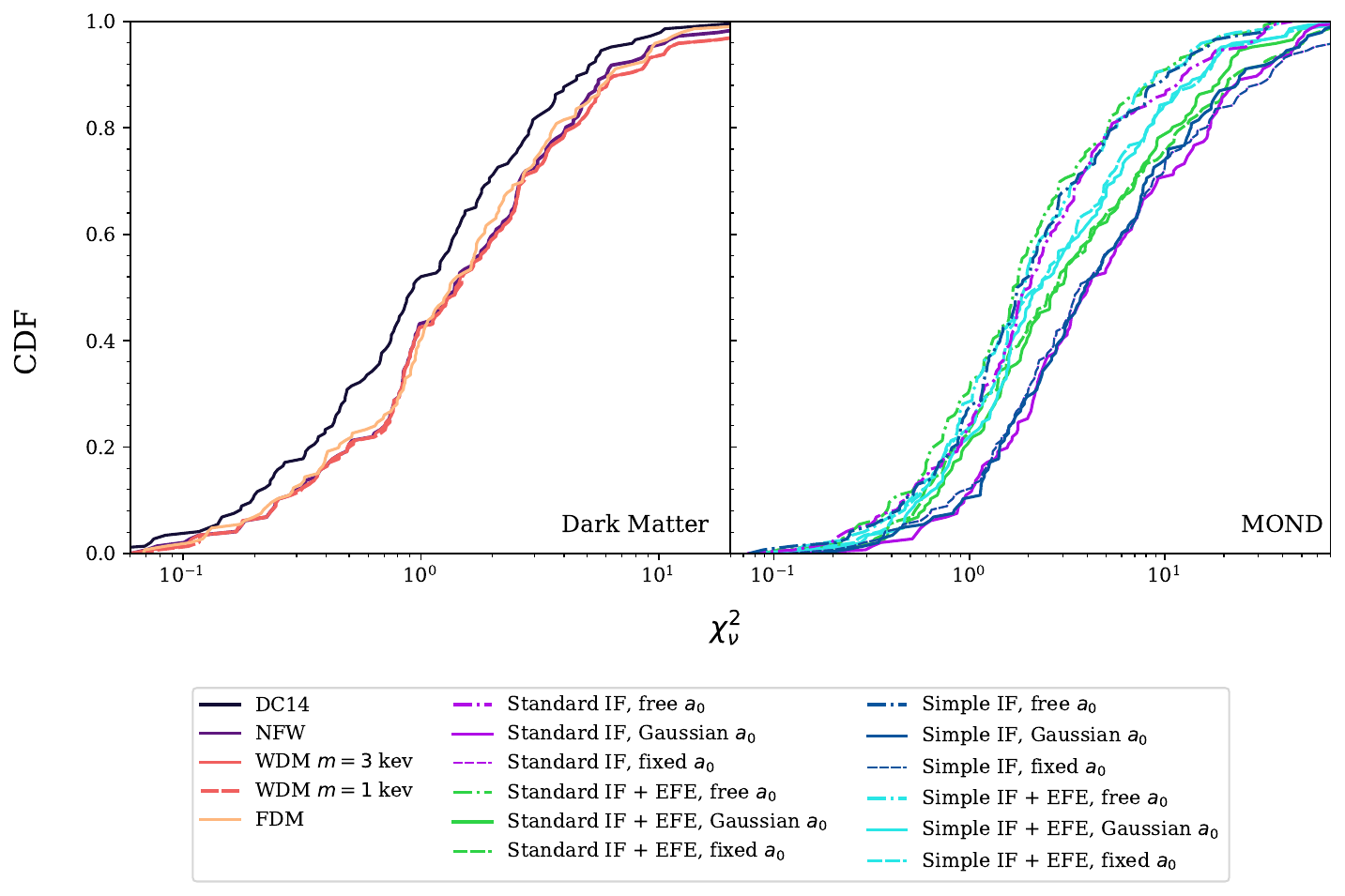}
        \caption{The cumulative distribution function (CDF) of the reduced chi-square statistic, $\chi^{2}_{\mathrm{\nu}}$, for each density profile over our $147$ galaxy subsample using the priors given in Table~\ref{table:free params}. \textit{Left Panel:} The $\chi^{2}_{\mathrm{\nu}}$-CDF of the 4 dark matter models. \textit{Right Panel:} The $\chi^{2}_{\mathrm{\nu}}$ of the 3 MOND interpolation functions assuming either a free, Gaussian or fixed $a_{0}$ prior. In these fits, the disc inclination and the galaxy distance are kept fixed to their fiducial values from the \texttt{SPARC} database and not treated as nuisance parameters. This is particularly important for MOND fits (see Section~\ref{sec:MONDfit}).}
    \label{fig:Chi2} 
\end{figure*}

Figure~\ref{fig:Chi2} plots the cumulative distribution function (CDF) of the reduced chi-square statistic, $\chi^{2}_{\mathrm{\nu}}$, of the rotation curve fits. In Figure~\ref{fig:Chi2} we plot the $\chi^{2}_{\mathrm{\nu}}$-CDF of the dark matter and MONDian model MCMC fits separately in the left and right panels respectively, with the MCMC priors listed in Table~\ref{table:free params}. The individual median $\chi^{2}_{\mathrm{\nu}}$ values of each model are listed in Table~\ref{table:Chi2vals} for comparison. Our aim in Figure~\ref{fig:Chi2} is not to directly compare the performance (via Bayesian statistics) of the dark matter and MONDian models, since each framework is characterised by a distinct set of free parameters. Consequently, statistical measures such as $\chi^{2}_{\mathrm{\nu}}$ and $\Delta\mathrm{BIC}$ are not directly comparable between the two classes of models, as they implicitly weigh all parameters equally despite their differing physical significance and influence on the RC fits. In dark matter models, the baryonic mass distribution generally plays a subdominant role because the halo contribution dominates the overall fit; accordingly, the halo parameters exert the greatest influence on the inferred rotation curves. By contrast, MONDian fits depend entirely on the baryonic mass distribution, rendering $\Upsilon_{*}$ substantially more important than in the dark matter case. Small inaccuracies in the baryonic component will therefore be much more relevant for MOND than for DM models. 
To prove these points, \citet{1998ApJ...508..132D} showed that fitting the rotation curve of a galaxy via the baryonic mass distribution of a different galaxy still retained a good fit with a DM halo, but failed using MOND. Following \citet{2017MNRAS.466.1648K}, in Figure~\ref{fig:Chi2} we do not include observational uncertainties in distance or inclination. However, as these effects may be non-negligible in certain models, particularly for MOND models \citep{2024A&A...689L...3L}, the median $\chi^{2}_{\mathrm{\nu}}$ for all models both with and without accounting for these observational uncertainties are reported in Table~\ref{table:Chi2vals}. A more detailed discussion on the comparison among different models is provided in Section~\ref{sec:Overfitting}.

\renewcommand{\arraystretch}{1.5}
\begin{table*}
\begin{tabular}{lcc|lcc}
\hline
\multicolumn{1}{|l|}{\textbf{Model}} & \multicolumn{1}{c|}{\textbf{No. of parameters}} & \multicolumn{1}{c|}{\textbf{Median $\chi^{2}_{\mathrm{\nu}}$}} & \multicolumn{1}{|l|}{\textbf{Model}} & \multicolumn{1}{c|}{\textbf{No. of parameters}} & \multicolumn{1}{c|}{\textbf{Median $\chi^{2}_{\mathrm{\nu}}$}} \\ \hline
& \multicolumn{2}{|c|}{\textbf{Fixed $D$ and $i$}}  & & \multicolumn{2}{|c|}{\textbf{Log-normal $D$ and $i$}}  \\ \hline
NFW  & 3  & 1.387 & NFW  & 5  & 1.485  \\
NFW $+$ \citet{Dutton2014} & 3  & 2.022 & - &&  \\
DC14 & 3 &  0.934  & DC14 & 5 &  1.157\\
DC14 $+$ \citet{Dutton2014} & 3 &  1.162   &  - && \\
WDM $m_{x} = 3 \:\mathrm{keV}$ & 3 &  1.401 &WDM $m_{x} = 3 \:\mathrm{keV}$ & 5 &  1.925 \\ 
WDM $m_{x} = 1 \:\mathrm{keV}$  & 3 & 1.440 & WDM $m_{x} = 1 \:\mathrm{keV}$ & 5 &  1.961\\
FDM & 4 &  1.324 &-&&\\
\hline

Standard IF, free $a_{0}$  & 2  & 2.030 & -&&\\
Standard IF, Gaussian $a_{0}$ & 2  & 4.019 & Standard IF, Gaussian $a_{0}$ & 4  & 2.988 \\
Standard IF, fixed $a_{0}$ & 1  & 3.667 &  Standard IF, fixed $a_{0}$ & 3  & 2.601\\

Simple IF, free $a_{0}$ & 2   & 1.756 & -&&\\

Simple IF, Gaussian $a_{0}$ & 2  & 3.888 & Simple IF, Gaussian $a_{0}$ & 4 & 2.680 \\
Simple IF, fixed $a_{0}$  & 1 & 3.667 & Simple IF, fixed $a_{0}$ & 3 & 2.435
\\
Standard IF + EFE, free $a_{0}$ & 3&  1.674 & -&&\\
Standard IF + EFE, Gaussian $a_{0}$  & 3  &  2.739 & Standard IF + EFE, Gaussian $a_{0}$ & 5 &  4.395 \\
Standard IF + EFE, fixed $a_{0}$ & 2 &  2.812 & Standard IF + EFE, fixed $a_{0}$  & 4  &  2.670 \\

Simple IF + EFE, free $a_{0}$ & 3&  1.926 & -&&\\
Simple IF + EFE, Gaussian $a_{0}$  & 3  &  2.335 & Simple IF + EFE, Gaussian $a_{0}$ & 5 &  2.467 \\
Simple IF + EFE, fixed $a_{0}$ & 2 &  2.237 & Simple IF + EFE, fixed $a_{0}$  & 4  &  2.197 \\

\hline                                        
\end{tabular}
\caption{Table of the median reduced chi-square statistic, $\chi^{2}$, values for dark matter and MOND models, using maximum likelihood parameter estimation. The NFW and DC14 model are included with and without the \citet{Dutton2014} $c_{\mathrm{vir}}-M_{\mathrm{vir}}$ relation as a prior on $c_{vir}$.}
    \label{table:Chi2vals}
\end{table*}
\renewcommand{\arraystretch}{1}

It must be noted that all of the median $\chi^{2}_{\mathrm{\nu}}$ values of the dark matter models listed in Table~\ref{table:Chi2vals}, except for the DC14 model, lie close to one another. The MOND models tend to exhibit larger median values of $\chi^{2}_{\mathrm{\nu}}$. However, as shown in the right columns of Table~\ref{table:Chi2vals}, the observational uncertainties in the galaxy distance and inclination play a significant role in improving the quality of MOND RC fits by roughly reducing the $\chi^{2}_{\mathrm{\nu}}$ by $\sim30\%$. On the other hand, accounting for these uncertainties in the dark matter model only moderately increases the median $\chi^{2}_{\mathrm{\nu}}$ due to the low reward of the additional two parameters. The WDM model becomes slightly more distinguishable from NFW and DC14 in this case, but not to a significant degree. The effect of introducing observational uncertainties in the analysis is discussed in more detail in Section~\ref{sec:Overfitting}. The median $\chi^{2}_{\mathrm{\nu}}$ of these models alone cannot conclusively distinguish whether one particular model is more successful than another, especially when accounting for the effect of additional cosmological priors. In the following subsections we discuss the median-$\chi^{2}_{\mathrm{\nu}}$ values of each individual model and discuss their impact.

In recent literature there have been some modifications to and discussion of the model put forward in \citet{Cintio_2014}, for example, the Dekel–Zhao (DZ) density profile \citep{2020MNRAS.499.2912F}, which is essentially a simpler, analytical adaptation to the DC14 profile. \citet{2020MNRAS.499.2912F} noted that the DZ profile fits the RCs of hydrodynamical simulations equally well to the DC14 profile so we do not expect a significantly different outcome switching to the DZ profile when fitting the data. \citet{2016MNRAS.456.3542T} suggested that although the DC14 is able to fit the DM cores, it tends to overestimate the outer regions. They favoured more flexible profiles such as the Einasto profile \citep{1965TrAlm...5...87E} for its ability to fit both the central and outer regions of RCs.

\subsubsection{NFW $\chi^{2}_{\mathrm{\nu}}$}

The NFW profile is also quite successful with a median $\chi^{2}_{\mathrm{\nu}} \sim1.4-1.5$. Although, on statistical grounds the NFW profile is an acceptable model, in several instances it still tends to overestimate the rotational velocity in the innermost radii. This is expected since the NFW profile is cuspy, hence the higher densities in the central regions tend to boost the rotational velocities beyond these measurements. It is known that the NFW profile well encapsulates the behaviour of the outer rotation curve of the galactic discs of spiral galaxies \citep{2007MNRAS.378...41S}. This is confirmed in our analysis of individual RCs, hence why many empirical cored density profiles are simply a modification of the inner section of the NFW profile, whilst keeping the NFW-envelope past the core region, such as the \texttt{CORE}-NFW profile \citep{2016MNRAS.459.2573R,2017ApJ...840...34S}. Other cored profiles such as the Einasto and Burkert profiles \citep{1965TrAlm...5...87E,1995ApJ...447L..25B} are shown to reproduce the RCs of spiral galaxies well \citep[e.g.,][]{2017MNRAS.470.2410R, 2019A&A...623A.123G,2020ApJS..247...31L}. In this work we chose to employ the DC14 model instead because it is physically motivated, based on the core production induced by baryonic feedback effects \citep[e.g.,][]{1994ApJ...427L...1F, 2011AJ....141..193O, 2012MNRAS.421.3464P}.

\subsubsection{DC14 $\chi^{2}_{\mathrm{\nu}}$}

 \begin{figure}
    \centering
    \includegraphics[width=0.475\textwidth]{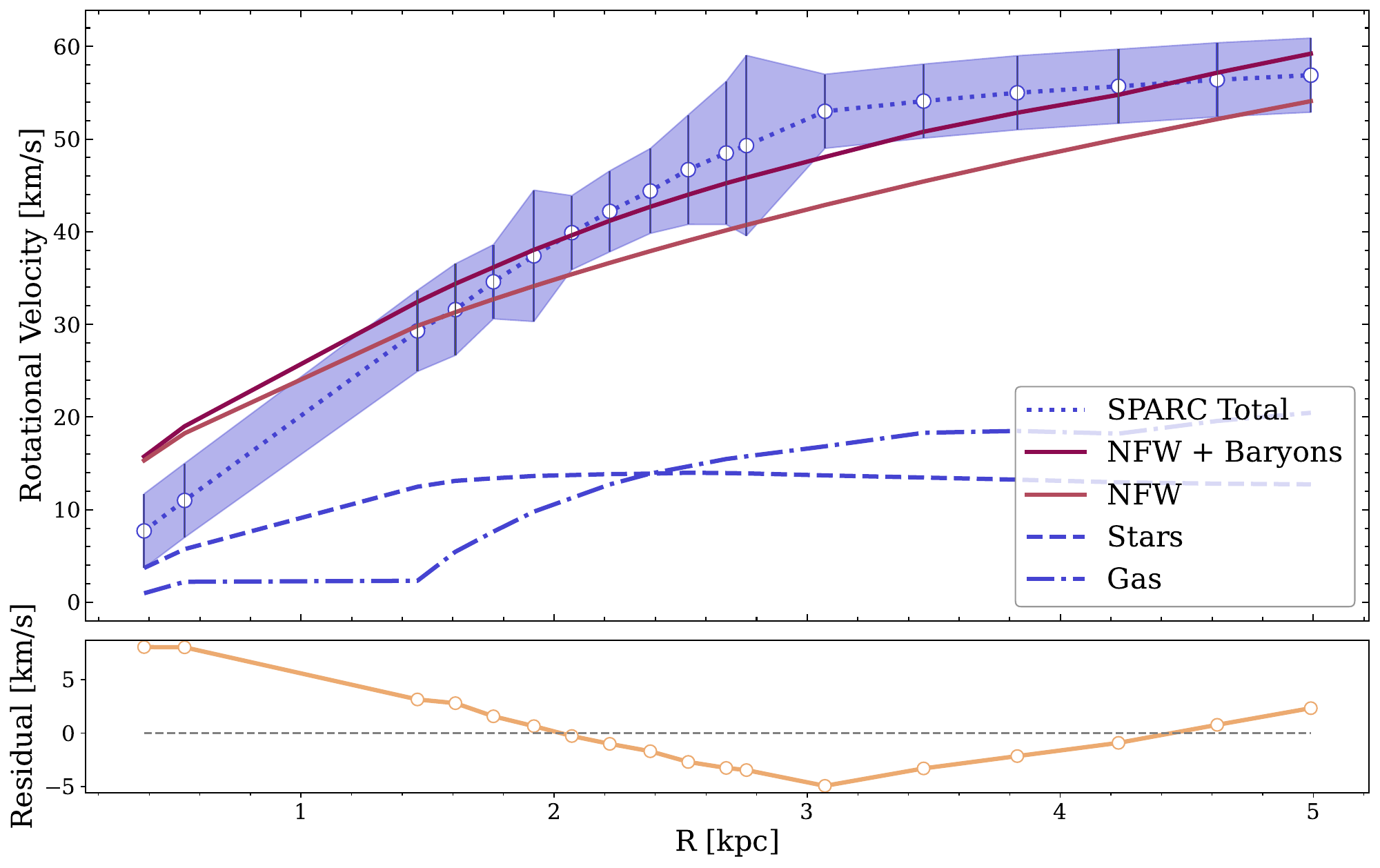}\hfill
    \includegraphics[width=0.475\textwidth]{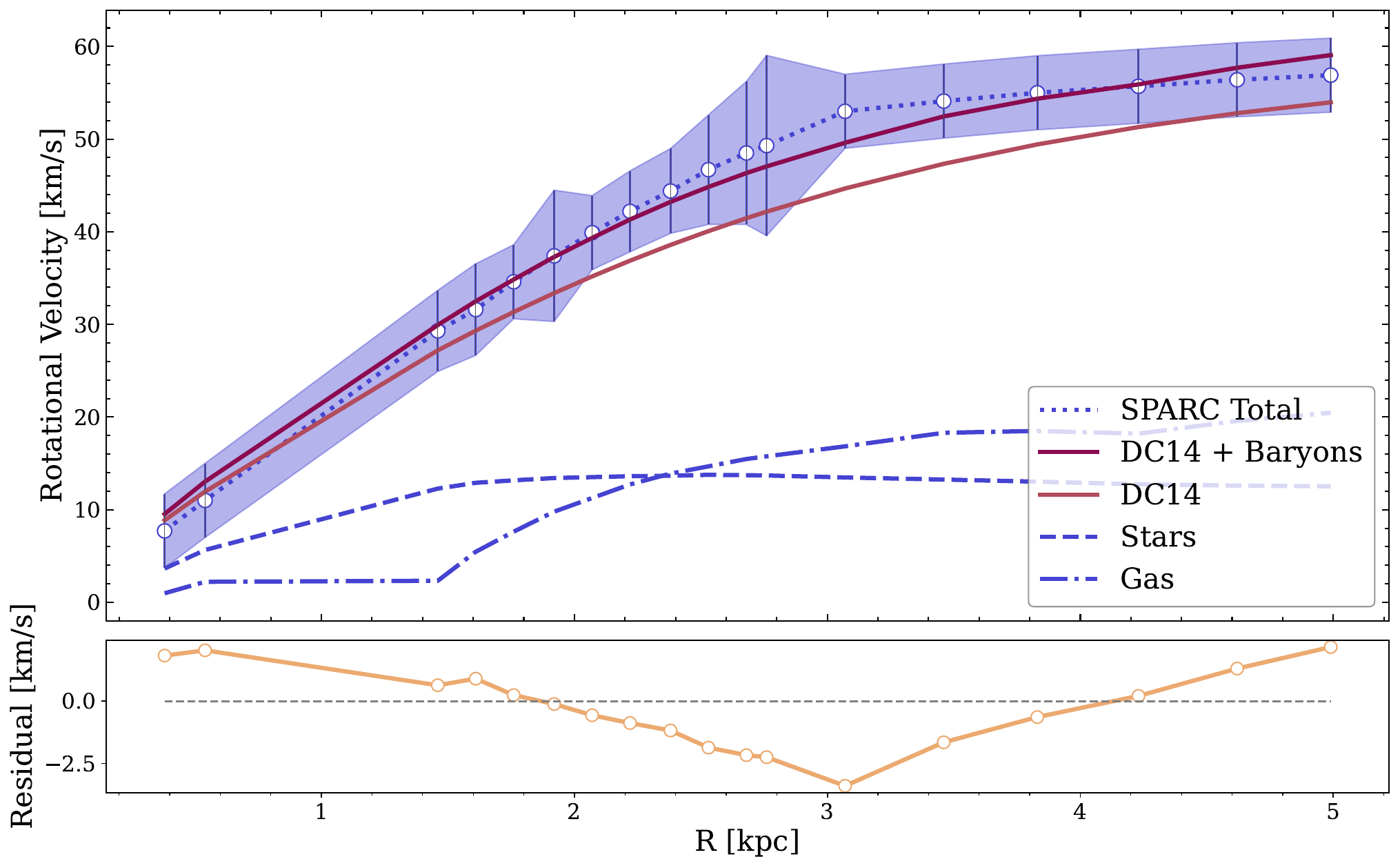}\hfill
        \caption{The rotation curve of UGC01281 assuming an \textit{Top Panel}: NFW dark matter profile and a \textit{Bottom Panel}: DC14 dark matter profile. The gas, stellar, and total velocity components from the \texttt{SPARC} database are represented by the blue dot-dashed line, dashed line, and data points, respectively. The blue shaded region denotes the $1\sigma$ uncertainty in the circular velocity measurements. The total best-fit velocity, shown by the red curve, is the sum of the stellar, gas, and dark matter (orange curve) components. In the bottom panel of each plot, the yellow curve represents the residuals between the observed and modelled total RCs.}
    \label{fig:UGC01281} 
\end{figure}

It is interesting to note from Table~\ref{table:Chi2vals} that the DC14 profile is only marginally better performing than NFW, with a median $\chi^{2}_\mathrm{{\nu}}$ of $\sim0.9$ against the $1.4$ from NFW, a difference that can be easily accounted for by measurement errors and/or RC inaccuracies \citep[e.g.,][]{2021A&A...646L..13L}. Indeed, when allowing for the distance and inclination to vary with Gaussian priors, the difference between DC14 and NFW is further reduced with mean values of $\chi^{2}_\mathrm{{\nu}}\sim1.2$ and $\sim~1.5$ respectively.   
Although on average their fitting power may not be too different, on a galaxy-by-galaxy basis, the performance of the DC14 and NFW can still be noticeably different. For example, in Figure~\ref{fig:UGC01281} we compare the rotation curve fits to UGC01281 for both the NFW and DC14 models, where the discrepancy between the inner halo profile and the total velocity in the NFW fit (top panel) is resolved by distributing matter outwards (bottom panel). Although UGC01281 is better fitted by the DC14 model, several other galaxies, especially those dominated in the inner regions by the stellar component, do not present a significant improvement when fitted with the DC14 model. For these galaxies it may imply that the maximum likelihood estimation of $\Upsilon_{*}$ could be overestimated, causing the stellar component to also be overestimated. In fact, by looking at the stellar components of these galaxies they overdominate the innermost regions and are sometimes equal to the total velocity. This issue highlights the degeneracy between the halo parameters (such as $M_{\mathrm{vir}}$ and $\Upsilon_{*}$ which both act to normalise the RCs) when fitting the large-scale dynamics. Interestingly, the performance of the DC14 (and NFW) model declines significantly when an $c_{\mathrm{vir}}-M_{\mathrm{vir}}$ prior is applied, with its $\chi^{2}_\mathrm{{\nu}}$ value rising. The prior significantly reduces the performance of both models, which have also been shown to not properly align with some of the halo scaling relations \citep{2017MNRAS.466.1648K,2020ApJS..247...31L}. Although the DC14 model aligns more closely with the \citet{Dutton2014} $c_{\mathrm{vir}}-M_{\mathrm{vir}}$ relation than the NFW profile, as shown in \citet{2017MNRAS.466.1648K}, many galaxies in the \texttt{SPARC} sample still fall outside the $2\sigma$ uncertainty range.

\subsubsection{WDM $\chi^{2}_{\mathrm{\nu}}$} \label{sec:WDM_Chi}

The WDM model performs similarly to the NFW profile. The $M_{\mathrm{vir}}$ inference of the NFW haloes and those of the WDM haloes are nearly identical across our sample. The thermal effects of a WDM model with a particle mass of $m_{\chi}=3$ keV does not have much of an impact on the halo parameter inference and this trend does not depend significantly on the chosen particle mass. When switching from 3 keV to 1 keV the $\chi^{2}_{\mathrm{\nu}}$ is in fact nearly unaltered, irrespective of whether we allow $D$ and $i$ to vary. (Table~\ref{table:Chi2vals}). A lighter $m_{\chi}$ leads to a lower halo concentration. Although the resulting halo remains cuspy rather than developing a constant-density core, its reduced central concentration leads to smaller rotational velocities in the inner region. The difference between the $m_{\chi}=1$ keV and $m_{\chi}=3$ keV rotation curve fits is too small to draw any firm conclusions as to which particle mass is preferred, as further discussed in Section~\ref{sec:WDM_dis}. In what follows, we focus on the $m_{\chi}=3$ keV case which is better aligned with the Lyman-$\alpha$ Forest measurements \citep{2005PhRvD..71f3534V,2023PhRvD.108b3502V}. It must be noted however, that lower masses such as 1 keV are not excluded, if the dark matter contributes to the total dark matter density by $<16\%$ \citep{2025PhRvD.112d3502G}.

\subsubsection{FDM $\chi^{2}_{\mathrm{\nu}}$}

The FDM halo consists of a solitonic core surrounded by an NFW-envelope. The FDM profile has an additional parameter compared to the NFW profile, namely the transition radius, $r_{t}$, between the solitonic core and the NFW envelope. Despite this increased flexibility, the FDM and NFW models end up with comparable fitting performance with $\chi^{2}_{\mathrm{\nu}}\sim1.3$ and $\chi^{2}_{\mathrm{\nu}}\sim1.4$, respectively. Overall, we find that the FDM model tends to fail in fitting the innermost portions of the RCs of several galaxies, indicating that the problem lies with the solitonic core. It performs well in the highest mass galaxies, where the solotonic region is relatively smaller compared to the full extent of the RC. From Equation~\ref{eq:sol} it can be seen that the solitonic core profile depends on the assumed value of the ultralight particle mass, $m_{\chi}$, which we assume to be $m_{\chi}=10^{-22}$ eV and constant across all galaxies. The continuity conditions cannot be fully achieved across our whole sample when assuming a fixed $m_{\chi}$. A constant value of $m_{\chi}$ for all galaxies may prevent a smooth density profile around the core radius, possibly suggesting the need for a family of ultralight particles of various masses to explain the vast diversity of rotation curve morphologies. This issue is highlighted in \citet{2023MNRAS.523.3393K}, who concluded that there is no single value of $m_{\chi}$ that satisfies the rotation curve diversity in the \texttt{SPARC} sample.

\subsubsection{MOND $\chi^{2}_{\mathrm{\nu}}$}\label{sec:MONDfit}

MOND compared to DM models tends to predict somewhat higher median reduced chi-square statistics, $\chi^{2}_{\mathrm{\nu}}$. However, as discussed previously, direct comparison of $\chi^{2}_{\mathrm{\nu}}$ between dark matter and MOND is of limited usefulness, since the corresponding free parameters fulfil fundamentally different roles and carry different levels of influence within each theoretical framework. As seen in Table~\ref{table:Chi2vals}, the incorporation of the External Field Effect (EFE) improves the quality of the RC fits by a significant amount for both the Simple and Standard IFs. In the case of the Standard MOND IF whilst assuming the theoretical value of $a_{0} = 1.2\times10^{-10}\:\mathrm{m\;s}^{-2}$, including the external acceleration ($a_{\mathrm{e}}$) as a free parameter reduces the $\chi^{2}_{\mathrm{\nu}}$ from $3.667$ to $2.812$, and similarly from $3.667$ to $2.237$ in the case of the Simple IF. This effect allows the RC to have more flexibility in the outer regions of these fits. A common feature among the MOND curves is the manifestation of Renzo’s rule, which arises due to the proportionality between the MONDian rotational velocity and the baryonic circular velocity. In galaxies where the baryonic contribution rises less steeply than the observed total velocity curve, MOND systematically underestimates the outer rotation curve amplitudes. As a consequence, if the distance and inclination are not treated as nuisance parameters, the MCMC fitting procedure tends to increase $\Upsilon{*}$ to elevate the overall rotation curve. While this adjustment improves the fit in the outer regions, it typically results in an overestimation of the inner velocity profile. The effect of $a_{\mathrm{e}}$ reducing the outer velocities of the total RCs can mediate this issue, thus resulting in a better fit to the data.
We find that both MOND interpolating functions, Simple and Standard, can generate acceptable fits to disc galaxy RCs in line with previous dedicated studies \citep[e.g.,][]{2005MNRAS.363..603F,2008ApJ...683..137M,2008ApJ...684.1018G,Gentile_2011}, however, there is a slight preference for the Simple MOND interpolation function. 

The MOND paradigm postulates a universal acceleration constant, $a_{0}$. As a quasi-reductio exercise, we initially adopt a flat prior on $a_{0}$ around $=[0.1,4]\times 10^{-10}\:\mathrm{ms^{-2}}$, which lowers the effective $\chi^{2}_{\mathrm{\nu}}$. We include this treatment of $a_{0}$ solely to demonstrate, as outlined by \citet{2021A&A...646L..13L}, that permitting fundamental constants to vary as free parameters within a Bayesian framework can lead to an artificial apparent disagreement with a universal constant value across the sample. Although allowing $a_{0}$ to vary as a free parameter can, at face value, improve rotation curve fits, this freedom introduces significant degeneracies with other model parameters, such as $\Upsilon_{*}$, distance, and inclination. As discussed by \citet{2021A&A...646L..13L}, these degeneracies can lead to apparent variations in $a_{0}$ that do not necessarily reflect a true physical departure from a universal value. Indeed, we also find, for example, that a Standard MOND with free $a_{0}$ and one with fixed $a_{0}$ but varying $D$ and $i$, yield comparable $\chi^{2}_{\mathrm{\nu}}$ of $\sim2.0$ and $\sim2.6$ respectively, and $\sim1.8$ and $\sim2.4$ for the Simple MOND, with a minimally larger value when $a_{0}$ is fixed but $D$ and $i$ vary.

\subsubsection{BIC Statistic}\label{sec:Overfitting}

Traditionally, in rotation curve fitting the reduced chi-square statistic is used to analyse a model's success. This statistic however, does not penalise models for having additional parameters, therefore limiting the effectiveness of the comparison among significantly diverse models. A more appropriate method of comparing models with differing numbers of parameters, such as those in this work, is the BIC statistic, which penalises additional complexity within a model. The BIC statistic is expressed as

\begin{equation}\label{BIC}
    \mathrm{BIC} = -2\ln(\mathcal{L})+p\ln(N)
\end{equation}
where $\mathcal{L}$ is the likelihood function from Equation~\ref{eq:log-like}, $p$ is the number of free parameters for a given model and $N$ is the number of measurements per galaxy rotation curve. As evident from Equation~\ref{BIC}, for models that equally well minimise the likelihood, the BIC disfavours those models characterised by a larger number of parameters for a given set of fitted data points. Therefore, the BIC statistic retains the same limitations inherent in the $\chi^{2}_{\mathrm{\nu}}$ statistic, where statistical and systematic uncertainties in the RCs can propagate and impact MOND more than DM models. As for the $\chi^{2}_{\mathrm{\nu}}$ tests, we thus continue comparing DM and MOND models separately, and mostly testing the performance of each variant within its own class of models. Specifically, we choose to compare all dark matter models to the DC14 profile (with and without a log-normal prior of $D$ and $i$) using the difference in their BIC statistics, given by

\begin{equation}\label{Delta BIC}
    \Delta\mathrm{BIC} = \mathrm{BIC}_{X}-\mathrm{BIC}_{\mathrm{DC14}}
\end{equation}
for any dark  model, X. In Table~\ref{table:BIC} we show the percentages of galaxies in our subsample that are preferred ($\Delta\mathrm{BIC}>2$), and strongly preferred ($\Delta\mathrm{BIC}>6$) by DC14 for each of the other models. 

\begin{figure*}
    \centering
    \includegraphics[width=\textwidth]{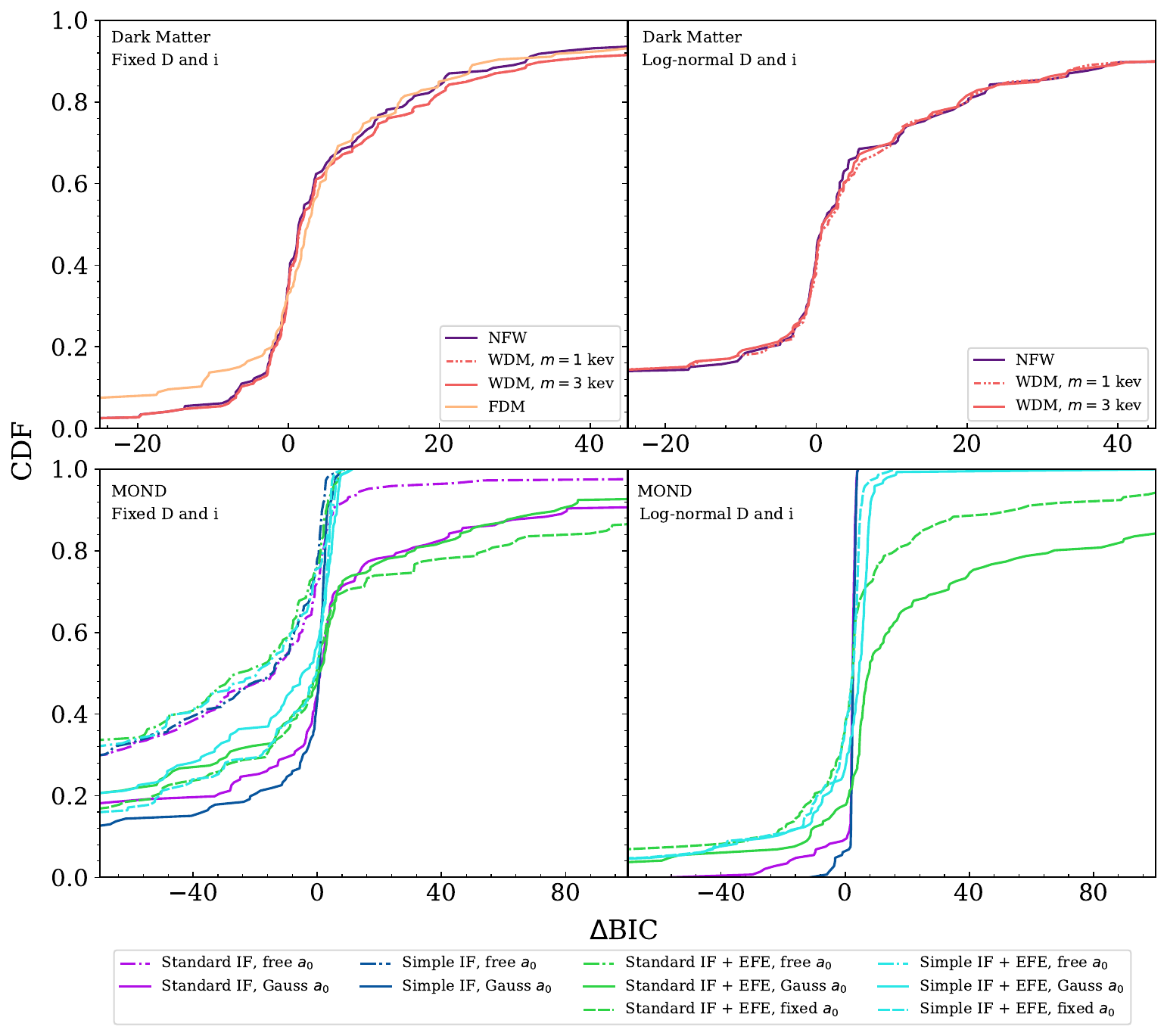}
    \caption{The cumulative distribution function (CDF) of the $\Delta$BIC statistic of our $147$ late type galaxy subsample. \textit{Upper left panel:} The $\Delta$BIC-CDF where the DC14 model is the reference model. All models here assume the measured distance and inclination as a fixed value. \textit{Upper right panel:} The $\Delta$BIC-CDF where the DC14 model (with a log-normal prior on distance and inclination) is the reference model. All models here assume a log-normal prior on distance and inclination. \textit{Lower left panel:} The $\Delta$BIC-CDF where the Standard/Simple MOND IFs with a fixed prior of $a_{0}=1.2\times 10^{-10} \:\mathrm{m\;s}^{-2}$ is the reference model. All models here assume the measured distance and inclination as a fixed value. \textit{Lower right panel:} The $\Delta$BIC-CDF where the Simple MOND IF with a fixed prior of $a_{0}$ (with a log-normal prior on distance and inclination) is the reference model. All models here assume a log-normal prior on distance and inclination. Refer to Table~\ref{table:BIC} for the percentages of the sample that prefer the reference model.}
    \label{fig:BIC}
\end{figure*}

\renewcommand{\arraystretch}{1.5}
\begin{table*}
\centering
\begin{tabular}{|l|c|c|l|c|c|c|}
\hline
\multicolumn{1}{|l|}{\textbf{Model (Fixed $D$ and $i$)}} &  & & {\textbf{Model (Log-normal $D$ and $i$)}} &  & \\ \hline
Dark Matter (All Galaxies) &\multicolumn{1}{c}{\textbf{$\Delta\mathrm{BIC}>2$}} &\multicolumn{1}{c|}{\textbf{$\Delta\mathrm{BIC}>6$}}  &  & \multicolumn{1}{c}{\textbf{$\Delta\mathrm{BIC}>2$}}& \multicolumn{1}{c|}{\textbf{$\Delta\mathrm{BIC}>6$}} \\ \hline
                     
NFW & 47.0$\%$ & 33.1$\%$   &  NFW & 46.5$\%$ & 31.4$\%$\\
WDM $m_{x} = 3 \:\mathrm{keV}$   & 48.4$\%$ & 34.4$\%$  & WDM $m_{x} = 3 \:\mathrm{keV}$  & 47.1$\%$ & 32.6$\%$\\
WDM $m_{x} = 1 \:\mathrm{keV}$    & 44.4$\%$ & 34.9$\%$ &WDM $m_{x} = 1 \:\mathrm{keV}$ & 48.2$\%$ & 34.6$\%$\\
FDM  & 55.2$\%$ & 34.1$\%$ &  \\ 

\hline

Dark Matter (dwarf-only, $\log_{10}(L/L_{\odot})<10$) & & &  & & \\ \hline

NFW & 58.5$\%$ & 40.1$\%$   &  NFW & 52.4$\%$ & 35.7$\%$\\
WDM $m_{x} = 3 \:\mathrm{keV}$   & 58.5$\%$ & 40.1$\%$  & WDM $m_{x} = 3 \:\mathrm{keV}$  & 52.2$\%$ & 35.4$\%$\\
WDM $m_{x} = 1 \:\mathrm{keV}$    & 58.6$\%$ & 41.0$\%$ &WDM $m_{x} = 1 \:\mathrm{keV}$ & 54.3$\%$ & 37.8$\%$\\
FDM  & 68.7$\%$ & 47.1$\%$ &  \\ 

\hline
Standard IF & & & & & \\ \hline

Standard IF, free $a_{0}$  & 19.5$\%$ & 9.10$\%$  & &&\\
Standard IF, Gaussian $a_{0}$  & 46.0$\%$ & 30.0$\%$ & Standard IF, Gaussian $a_{0}$  &  79.6$\%$& 0.00$\%$\\
Standard IF + EFE, free $a_{0}$   & 15.8$\%$ & 3.44$\%$ &&\\
Standard IF + EFE, Gaussian $a_{0}$   & 47.4$\%$ & 31.3$\%$&Standard IF + EFE, Gaussian $a_{0}$   & 78.3$\%$ & 57.0$\%$\\
Standard IF + EFE, fixed $a_{0}$& 42.1$\%$ & 32.2$\%$ &Standard IF + EFE, fixed $a_{0}$ & 53.4$\%$ & 29.8$\%$\\

\hline
Simple IF & & & & & \\ \hline

Simple IF, free $a_{0}$  & 8.15$\%$ & 0.01$\%$ &\\
Simple IF, Gaussian $a_{0}$   & 27.5$\%$ & 1.93$\%$ & Simple IF, Gaussian $a_{0}$ & 79.6$\%$ & 0.00$\%$\\
Simple IF + EFE, free $a_{0}$   &  19.8$\%$& 3.90$\%$ &&&\\
Simple IF + EFE, Gaussian $a_{0}$   &33.9$\%$& 11.2$\%$ & Simple IF + EFE, Gaussian $a_{0}$   & 65.5$\%$ & 35.8$\%$\\ 
Simple IF + EFE, fixed $a_{0}$ & 34.6$\%$& 0.10$\%$ & Simple IF + EFE, fixed $a_{0}$   & 52.5$\%$ & 4.92$\%$\\

\hline                                        
\end{tabular}
\caption{Table of the $\Delta$BIC statistics compared to the DC14 model (top two panels) and the Standard/Simple IF with a fixed prior of $a_{0}=1.2\times 10^{-10} \:\mathrm{m\;s}^{-2}$ (third and fourth panels respectively) with the distance and inclination of each galaxy fixed (left columns) at their measured value and the distance and inclination as a log-normal prior (right columns) around their measured value. The second and third columns of each left and right sections present the percentage of galaxies in the sample that prefer, and strongly prefer the reference model over each other model of this work. The DM models are split into comparisons of the full $147$ galaxy sample (first panel), and the dwarf-only subsample (second panel).}
    \label{table:BIC}
\end{table*}
\renewcommand{\arraystretch}{1}

We then repeat this analysis using the Standard and Simple MOND IFs, adopting a fixed prior of $a_{0}=1.2\times 10^{-10} \:\mathrm{m\;s}^{-2}$ (as prescribed by the original MOND proposition) as the reference model for computing $\Delta\mathrm{BIC}$ for all MONDian variants. This choice ensures that all $\Delta$BIC values are interpreted relative to a physically consistent MOND baseline. All Standard IF variants are only compared to the Standard IF with fixed $a_{0}$, and similarly all Simple IF variants are compared to the Simple IF with fixed $a_{0}$.

In Table~\ref{table:BIC} we present the $\Delta$BIC statistics of dark matter (top two panels) and MOND (bottom two panels) separately, with (right columns) and without (left columns) a log-normal prior on $D$ and $i$. Looking at the summary statistics from the left column where $D$ and $i$ are held fixed, we note that in the case of the NFW model, there is a $47.0\%$ moderate preference, and a $33.1\%$ strong preference for the DC14 profile. For comparison, \citet{2017MNRAS.466.1648K} concluded that, when using their \texttt{SPARC} subsample (identical to ours), $~62\%$ and $~46\%$ of the sample prefer and strongly prefer DC14, larger fractions than our more stringent ones, most probably induced by the slight differences in prior spaces. We suggest that the remaining fraction of galaxies showing no preference for either DC14 or NFW are likely those with small or no cores, where model degeneracies arise due to closely similar likelihoods across the competing fits. We also find that the WDM according to the BIC statistic performs very similarly to the NFW, as expected since the particle masses of $1$ and $3$ keV typically induce only a mild mass redistribution in the profile at these mass scales. The FDM model (with $55.2\%$ and $34.1\%$ of the sample preferring and strongly preferring DC14) has a BIC distribution similar to NFW, despite the extra parameter, $r_{\mathrm{t}}$. When including the uncertainties around the measured distances and inclinations of each galaxy, via the use of log-normal priors, the NFW and DC14 comparison is unchanged within uncertainties, whereas the WDM models only marginally improve. We conclude that when comparing dark matter models, accounting for the measurement uncertainties in $D$ and $i$ does not aid in distinguishing one model from another, but rather strengthens their degeneracies.

We also calculate the BIC statistics of a dwarf-only subsample defined by $79$ galaxies with luminosities lower than $10^{10}\,L_{\odot}$, where differences between cuspy and cored halo profiles should become more evident. We find that there is a stronger preference for the DC14 model in the dwarf-only sample than in the full sample of $147$ \texttt{SPARC} galaxies, highlighting the improved flexibility of DC14 in fitting the more cored profiles of dwarf galaxies. In the dwarf-only sample the degeneracies between the other dark matter models only gets worse, with NFW and WDM being indistinguishable. This trend persists even when $D$ and $i$ are assigned log-normal priors. In the dwarf-only subsample the DC14 is noticeably favoured over FDM when a particle mass of $m_{\chi}=10^{-22}\,\mathrm{eV}$ is assumed, with a strong preference for DC14 in $\sim47\%$ of the subsample.

For the MOND comparison, we adopt the Standard and Simple IF with a fixed prior of $a_{0}=1.2\times 10^{-10} \:\mathrm{m\;s}^{-2}$ as the reference models. Although allowing $a_{0}$ to vary freely produces lower median BIC values, such models do not constitute faithful realisations of MOND, and are therefore not considered appropriate for comparison within the framework of the theory. We include the statistics of the MOND models with a free $a_{0}$ for completeness, and to demonstrate the degeneracies among the parameters used in this work. When comparing the performance of MOND with fixed and Gaussian priors on $a_{0}$, the results do not indicate a uniform preference for either the reference model or a Gaussian prior on $a_{0}$. In the Standard IF case, the fixed $a_{0}$ model is preferred for $46\%$ of the sample, whereas $\sim30\%$ of the sample prefer the Gaussian prior on $a_{0}$. We see a similar case for the Simple IF, where the fixed prior is preferred by $\sim28\%$ of the sample, and the Gaussian prior is preferred by $\sim30\%$. For the Standard IF + EFE model, there is a preference for the reference model in $\sim42\%$ of the sample when $a_{0}$ is fixed, rising to $\sim47\%$ when a Gaussian prior on $a_{0}$ is adopted, with strong preferences ($\Delta$BIC$>6$) in $\sim32\%$ and $\sim31\%$ of the sample, respectively. This behaviour is likely attributable to the limited improvement in fit provided by introducing $a_{\mathrm{e}}$ as an additional free parameter in some galaxies, rather than constituting evidence against the physical existence of the EFE. However, there is also an approximately $\sim40-50\%$ strong preference ($\Delta$BIC$<-6$) for the Standard and Simple IF + EFE models with both a fixed and Gaussian $a_{0}$ prior over the reference model.

The relative rigidity of the MOND rotation curve fits arises from the requirement that the acceleration scale, $a_{0}$, remains fixed or tightly constrained by the theory, limiting the ability of the model to fully reproduce the diverse morphologies of measured RCs. Allowing the observed distances, $D$, and inclinations, $i$, to vary within their observational uncertainties provides additional freedom to accommodate these discrepancies, thereby improving the overall quality of the fits. In the bottom right panel of Figure~\ref{fig:BIC}, and two bottom right panels of Table~\ref{table:BIC}, we report the results of the statistical analysis when including log-normal priors around distance and inclination. The reference model is favoured over the Standard IF with a Gaussian $a_{0}$ prior for $79.6\%$ of the sample, with a strong preference ($\Delta$BIC$>6$) in none of the galaxies. Thus, despite the high fraction of galaxies showing a moderate preference for the reference model, there is no strong BIC evidence against the Gaussian $a_{0}$ model in this case. For the Standard IF + EFE model, the reference model is favoured in $78.3\%$ of galaxies when a Gaussian prior on $a_{0}$ is adopted, with a strong preference in $57.0\%$, compared with $53.4\%$ and $29.8\%$, respectively, when $a_{0}$ is fixed. This indicates that the additional flexibility here does not add to the benefit of the fitting procedure when $D$ and $i$ uncertainties are accounted for. Also we highlight that the addition of the EFE does not improve the fits as much as it does when $D$ and $i$ are fixed. Allowing the distances and inclinations to vary therefore reduces the importance of the EFE parameters and highlights that the fixed $a_{0}$ case performs better than the Gaussian one.

\subsection{Luminosity-Halo Mass Relation} \label{LHMR}

In this section we present the Luminosity-Halo-Mass Relation (LHMR) as an additional and independent probe of the predictive power of the DM models considered in this work. The LHMR is a relation between the absolute $3.6\mathrm{\mu m}$-band luminosity, and the virial halo mass for CDM, WDM, and FDM. 

Here, at variance with \citet{2017MNRAS.466.1648K}, we choose to focus on the correlation between luminosity, $L$, and halo mass, $M_{\mathrm{vir}}$, as $L$ is a directly measurable quantity and does not rely on the additional mass-to-light ratio, $\Upsilon_{*}$ parameter. When comparing rotation curve model outputs with any $M_{*}-M_{\mathrm{vir}}$ from the literature (see references above), several systematics can arise in terms of the differences in $\Upsilon_{*}$ when calibrating stellar masses in different bands and under different assumptions between SPARC and the stellar mass functions used in the various studies. Here we instead rely on directly comparing the \texttt{SPARC} absolute $3.6\mathrm{\mu m}$-band luminosity and the inferred halo masses with the LHMR derived from abundance matching in the same waveband.

We derived a LHMR using the halo abundance matching (HAM) method between the $3.6\mathrm{\mu m}$-band galaxy luminosity function (LF) from \citet{2009ApJ...697..506D}, and the halo mass function (HMF) from \citet{2008ApJ...688..709T}, to directly compare model predictions from fitting single sources to the theoretically predicted halo masses, $M_{\mathrm{vir}}$, from our MCMC analysis. It must be noted that the \texttt{SPARC} luminosities and \citet{2009ApJ...697..506D} LF data are both given in the Vega magnitude system. The $3.6\mathrm{\mu m}$ band LF from  \citet{2009ApJ...697..506D} has a completeness of $92\%$ which does not differ significantly with magnitude, so should not cause any bias on the abundance matching procedure. We calculate the \citet{2008ApJ...688..709T} HMF using the \texttt{COLOSSUS} Python package \citep{2018ApJS..239...35D} for the CDM case corrected for the contribution of surviving subhaloes as described in Section~\ref{sec:HAM}. We use the redshift-dependent virial mass overdensity definition \citep{1998ApJ...495...80B} here in order to directly compare to the inferred virial halo masses from our maximum likelihood estimation.

\begin{figure*}
  \centering
  \includegraphics[width=\linewidth]{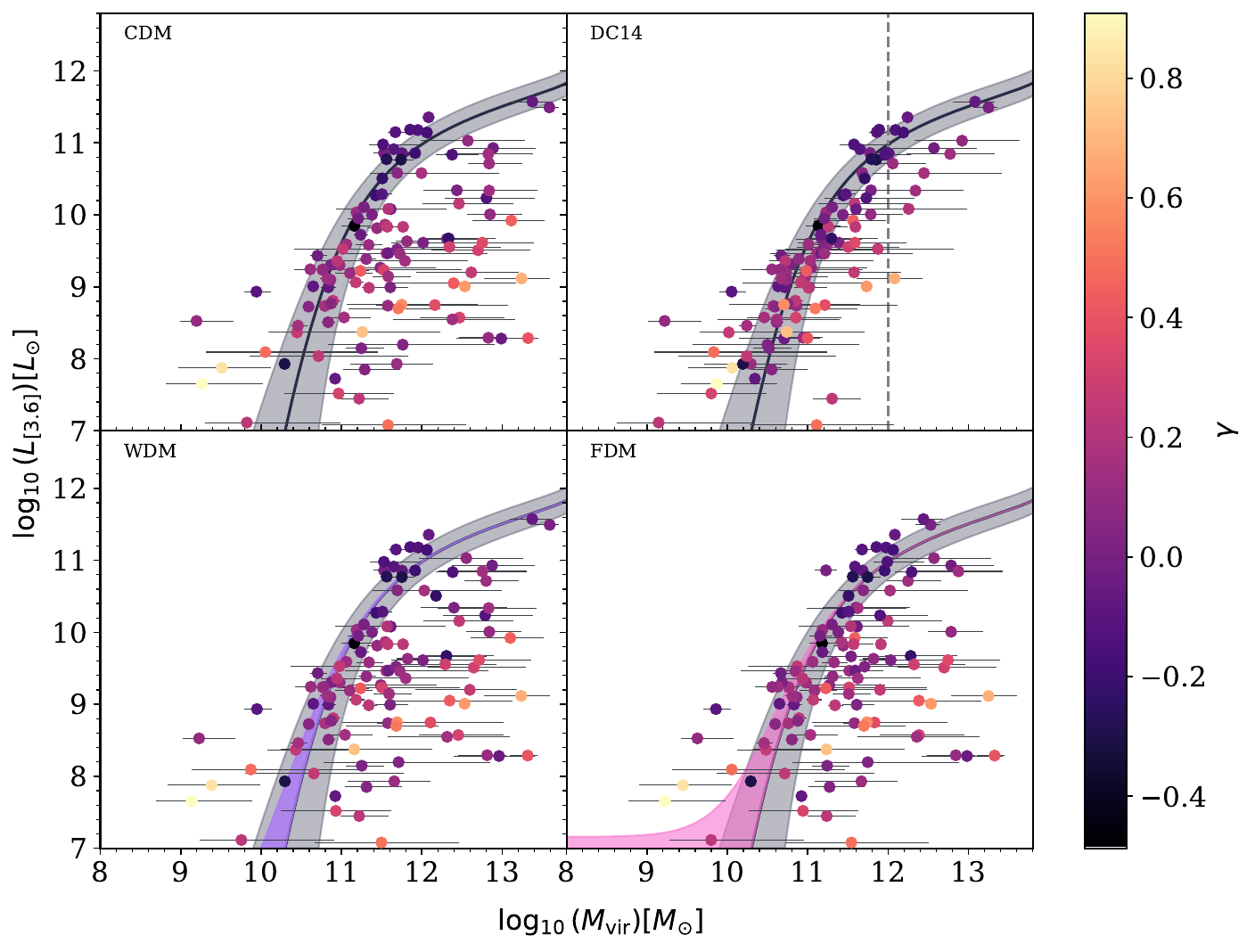}
  \caption{The Luminosity-Halo-Mass Relation (LHMR) resulting from the Halo Abundance Matching (HAM) technique between the HMF from \citet{2008ApJ...688..709T} and the $3.6\mathrm{\mu m}$-band LF from \citet{2009ApJ...697..506D}, compared with the maximum likelihood $M_{\mathrm{vir}}$ estimates from each density profile and observed $L_{[3.6]}$ values of each galaxy (scatter points). The colour bar indicates the logarithmic slope of the outer $30\%$ of each galaxy's rotation curve. The error bars show the $1\sigma$ error on the luminosities (too small to see) and the logarithmic of the virial halo mass, $M_{\mathrm{vir}}$. The black curve, purple, and pink shaded regions denote the expected LHMR assuming a CDM, WDM, and FDM Halo Mass Function (HMF) respectively. \textit{Top Left}: NFW profile. \textit{Top Right}: DC14 profile. \textit{Bottom Left}: WDM profile with $m_{\chi}=3$ keV. The purple shaded region denotes the LHMR for WDM with $m_{\chi}=[1-10]$ keV, with the lower particle mass resulting in the most deviation from the CDM curve. \textit{Bottom Right}: FDM profile. The pink shaded region denotes the LHMR for FDM with $m_{\chi}=[1-100] \times 10^{-22}$ eV, with the lower particle mass resulting in a more pronounced deviation from the CDM curve.}
  \label{fig:LHMR}
\end{figure*}

The LHMR curves resulting from abundance matching for each model are plotted in Figure~\ref{fig:LHMR}. The solid black line is the LHMR assuming the $\Lambda$CDM HMF, and the grey shaded region encompasses the bootstrapped $1\sigma$ errors on the measured LF originating from the uncertainty region. The purple shaded region in the bottom left panel represents the LHMR for WDM with $m_{\chi}=[1-10]$ keV, and the pink shaded region in the bottom right panel represents the LHMR for FDM with $m_{\chi}=[1-100] \times 10^{-22}$ eV, with the lower particle masses both resulting in a marked deviation from the CDM curve. The coloured data points are the $3.6\mathrm{\mu m}$-band absolute luminosities from the \texttt{SPARC} database against the maximum likelihood $M_{\mathrm{vir}}$ estimations. The colour bar labels the value of the logarithmic slope of the outer $30\%$ of each galaxy's RC, given by $\gamma=d\ln(v)/d\ln(r)$. It is interesting to note from Figure~\ref{fig:LHMR} that many of the galaxies departing from the predictions of the LHMR have non-flat, mostly rising RCs with $\gamma>0.1$, for which the determination of the host halo mass would be biased towards larger values. We quantify the performance of each dark matter model in reproducing the LHMR via the $\chi^{2}_{\mathrm{\nu}}$ listed in Table~\ref{tab:chi_LHMR}. The DC14 is the model that best reproduces the scatter around the mean relation, with a $\chi^{2}_{\mathrm{\nu}}=1.71$. The NFW and WDM, as expected, result in indistinguishable scatters, with $\chi^{2}_{\mathrm{\nu}}=7.87$ and $\chi^{2}_{\mathrm{\nu}}=7.89$ respectively. These two models predict also very similar distributions in the luminosity-halo mass plane for a particle mass of $m_{\chi}=3$ keV. We checked that a WDM model with $m_{\chi}=1$ keV yields essentially the same $L$-$M_{\mathrm{vir}}$ distribution, with $\chi^{2}_{\mathrm{\nu}}=7.60$. There is a slight systematic preference for larger haloes in \texttt{SPARC} compared to the HAM prediction, in agreement with \citet{2017MNRAS.466.1648K}. \citet{2019A&A...626A..56P,2019A&A...629A..59P} find a preference for smaller haloes at a fixed stellar mass, this is attributed to the fact that they impose the \citet{Dutton2014} concentration relation as a prior on $c_{\mathrm{vir}}$, whereas this work and \citet{2017MNRAS.466.1648K} find larger halo masses than HAM predicts when the concentration is prescribed a uniform prior between $1<c_{\mathrm{vir}}<100$. \citet{2017MNRAS.466.1648K} explore the effects of imposing the \citet{Dutton2014} concentration prior on the halo mass inference, and find better agreement with the \citet{2013MNRAS.428.3121M} SMHM relation, as found in \citet{2019A&A...626A..56P,2019A&A...629A..59P}. There may also be a possible shortcoming in the abundance matching procedure adopted here, requiring either circular velocity for the abundance matching \citep{2006ApJ...647..201C} or a different shape for the stellar mass function \citep{2019MNRAS.483.2506G}, as further discussed in Section~\ref{sec:Discussion}. We will explore some of these points in future work. \citet{2017MNRAS.466.1648K} and \citet{2020ApJS..247...31L} both also find that the NFW density profile produces an $M_{*}-M_{\mathrm{vir}}$ relation much broader than the one predicted by abundance matching. When the abundance matching relation is not given as a prior, the predicted distribution of halo masses at fixed galaxy mass/luminosity is large, particularly for the NFW profile, while the DC14, inclusive of baryonic feedback, shows the smallest halo scatter, in line with our results.
When looking at the individual RCs of the galaxies with luminosities $L<10^{10} \; \mathrm{L_{\odot}}$ and halo masses $M_{vir}>10^{12} \; \mathrm{ M_{\odot}}$, a clear trend of the velocity curve being too shallow tends to emerge. When adding a solitonic core to the galaxies as in the FDM model, the halo mass decreases accordingly due to the core steepening the rotation curve. This is likely the main driver as to why the FDM model predicts slightly less overly massive haloes than the NFW model and provides a slightly smaller scatter around the LHMR ($\chi^{2}_{\mathrm{\nu}}=6.20$). By applying the same cut in halo mass and luminosity to remove the large mass outliers of the FDM haloes, we find only $8$ galaxies in this group, as opposed to $17$ outliers when using the NFW profile. We do not show the DC14 outliers here as there is only $1$ galaxy that meets these conditions.

\begin{figure}
  \includegraphics[width=\linewidth]{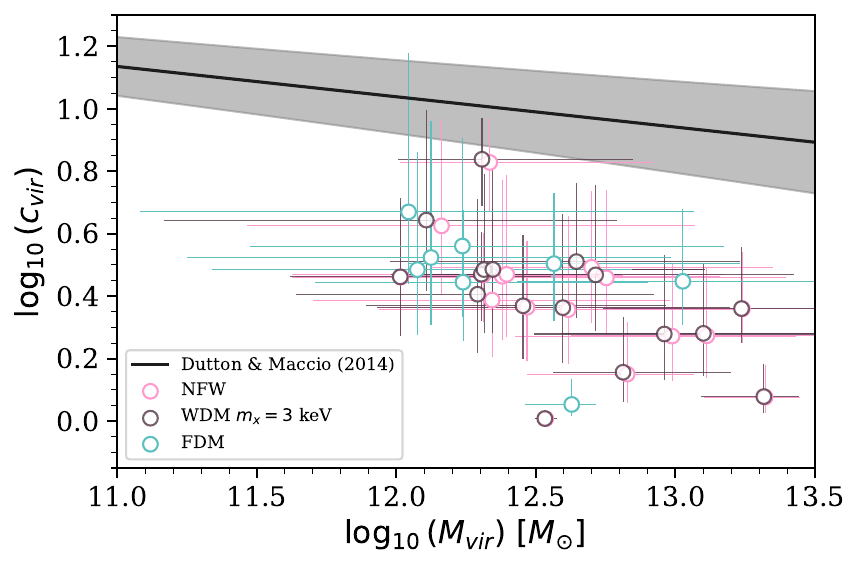}\hfill
  \caption{The NFW (pink), WDM (plum), and FDM (blue) best-fit values of the concentration and halo mass against the \citet{Dutton2014} $c_{\mathrm{vir}}-M_{\mathrm{vir}}$ relation (black line) with its $1\sigma$ scatter region (grey area). The scatter points show the values of the concentration and halo mass for the NFW, WDM, and FDM profiles with their corresponding 1$\sigma$ error bars.}
  \label{fig:Outliers}
\end{figure}

\begin{figure}
    \centering
    \includegraphics[width=0.475\textwidth]{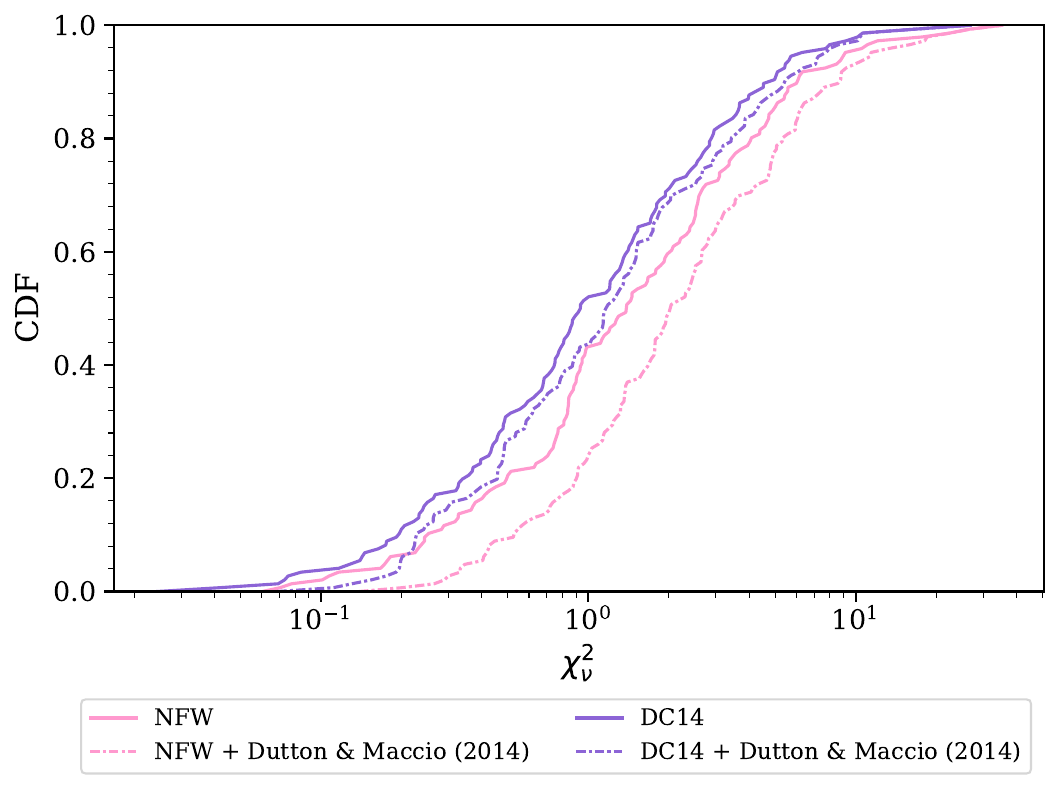}
        \caption{The cumulative distribution function (CDF) of the reduced chi-square statistic, $\chi^{2}_{\mathrm{\nu}}$, for the NFW (pink curves) and the DC14 (purple curves) profile over our $147$ galaxy subsample with (dot-dashed curves) and without (solid curves) a concentration prior around the \citet{Dutton2014} $c_{\mathrm{vir}}-M_{\mathrm{vir}}$ relation.}
    \label{fig:Chi2_DuttonPrior} 
\end{figure}

Our best-fit solutions to the RCs are also characterised by lower concentrations at fixed halo mass/virial radius than the typical values predicted by N-body simulations \citep[e.g.,][]{Navarro_1996, 2008MNRAS.390L..64D,  2011ApJ...740..102K,  2013MNRAS.436..503A, Dutton2014, 2016MNRAS.457.4340K, 2020ApJS..247...31L}, as shown in Figure~\ref{fig:Outliers}. By lowering the concentration, in fact, the inner scale radius tends to be larger, creating a more extended and less dense core. One solution is to include a prior around the \citet{Dutton2014} relation during the MCMC process. By doing so it can be seen in Figure~\ref{fig:Chi2_DuttonPrior} that the addition of these priors actually has a significantly negative effect on the fitting ability of the NFW model but a negligible impact on the predictions of the DC14 model, in agreement with the findings of \citet{2017MNRAS.466.1648K}. 

\begin{table}
    \centering
    \begin{tabular}{|c c c c c c|}
        \hline
        Model & NFW & DC14 & WDM, $m_{\chi}=3$ keV & FDM  \\
        \hline  
        $\chi^{2}_{\mathrm{\nu}}$ & 7.87 & 1.71 & 7.89 & 6.20 \\ 
        \hline
    \end{tabular}
    \caption{The reduced chi-square statistic, $\chi^{2}_{\mathrm{\nu}}$ of the halo masses of each dark matter model compared to the abundance matched Luminosity-Halo-Mass Relation.}
    \label{tab:chi_LHMR}
\end{table}

\subsection{Mean Rotation Curves up to 1 Mpc} \label{sec:MeanRC}

In the previous sections we compared different DM and MOND models against the individual \texttt{SPARC} RCs. However, this exercise is limited to a few tens of kiloparsecs. Here we extend the analysis to include the recent weak lensing results by \citet{2024ApJ...969L...3M} which measured RCs up to 1 Mpc. They used weak lensing measurements around isolated galaxies from the \texttt{KiDS} survey, employing a de-projection method that calculates the radial gravitational acceleration profiles from stacked excess surface density profiles, to test the predictions of gravitational potentials at larger distances. Their findings suggest that dark matter haloes may have more extended and flatter density profiles than those described by the single-halo NFW model. These observations could indicate the need for modifications to the NFW model or even alternative gravity or modified inertia models. 

To test the above argument we combine the rotation curve data from our \texttt{SPARC} sample of galaxies with the mean weak lensing data from \citet{2024ApJ...969L...3M}. We focus on each of the baryonic mass bin collected by \citet{2024ApJ...969L...3M}, namely $\log_{10}(M_{\mathrm{b}}/\mathrm{M_{\odot}}) = \left[9.0, 10.5\right],\;\left[10.5, 10.8\right],\;\left[10.8, 11.1\right],\;\mathrm{and}\;\left[11.1, 11.5\right]$, which, from now on, we will refer to as mass bins 1-4, respectively. We calculate which individual \texttt{SPARC} galaxies fall into this baryonic mass bin assuming an average $\Upsilon_{*}=0.5\;\mathrm{M_{\odot}}/\mathrm{L_{\odot}}$ as in \citet{2024ApJ...969L...3M}. The baryonic mass is then calculated as

\begin{equation}
    M_{\mathrm{b}} =\Upsilon_{*} L_{\mathrm{tot}} + 1.33 M_{\mathrm{HI}}.
\end{equation}
where the factor of $1.33$ accounts for the Helium contribution to the baryonic mass. 

\begin{figure*} 
  \centering
  \includegraphics[width=0.9\linewidth]{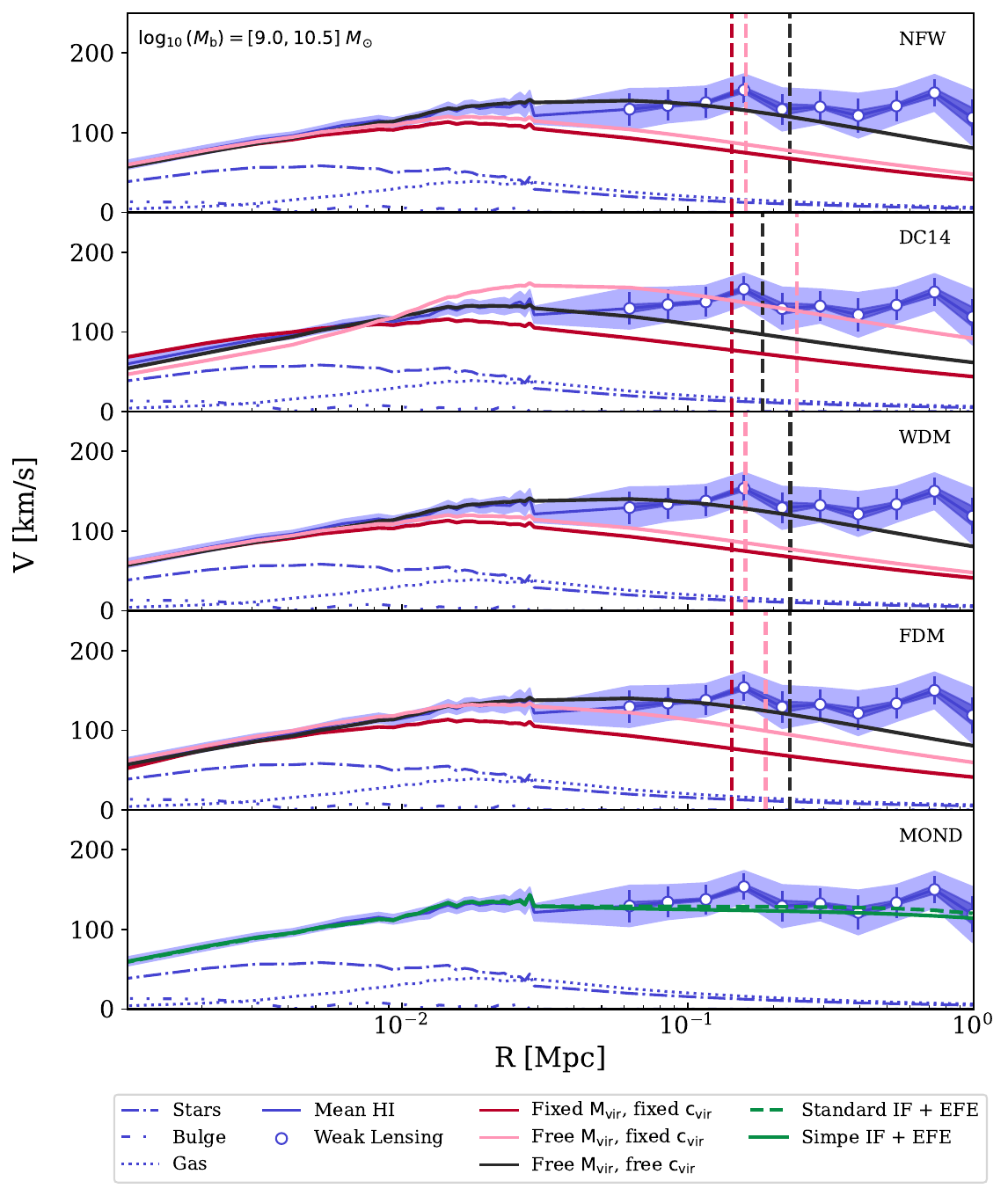}\hfill
  \caption{The best-fit total RCs of each dark matter model, and MOND compared with the mean HI and weak lensing data for galaxies in the range $\log_{10}(M_{\mathrm{b}}/\mathrm{M_{\odot}}) = [9.0,10.5]$. The red curves show the model RCs using halo parameters from \citet{Dutton2014} and the LHMR (Figure~\ref{fig:LHMR}), assuming the mean luminosity of the \texttt{SPARC} galaxies in this baryonic mass bin. The pink solid curves show the RCs obtained when $M_{\mathrm{vir}}$ is free and $c_{\mathrm{vir}}$ is fixed by the $c_{\mathrm{vir}}-M_{\mathrm{vir}}$ relation. The black curves show the best fits with both $M_{\mathrm{vir}}$ and $c_{\mathrm{vir}}$ as free parameters. The green solid and dashed curves in the bottom panel show the best-fit RCs assuming the Simple and Standard IF + EFE, with a fixed $a_{0}$ respectively. Vertical dashed lines mark the corresponding $R_{\mathrm{vir}}$. The HI and weak lensing data given by the blue solid curve and the scatter points respectively. The blue shaded region around the HI data represents the error on the mean velocities within each radial bin. The error bar of the scatter points show the statistical errors on the weak lensing data, the dark blue shaded region shows the systematic uncertainties, and the light blue shaded region around the weak lensing data shows the summation of both the statistical and systematic uncertainties. The dash-dotted, dotted, and dashed blue curves show the mean stellar, gas, and bulge component velocity curves respectively. Beyond the peak mean stellar and gas velocity we assume a $r^{-1/2}$ drop-off.}

  \label{fig:Mean_RC_Best_Fit_1}
\end{figure*}

\begin{figure*} 
  \centering
  \includegraphics[width=0.95\linewidth]{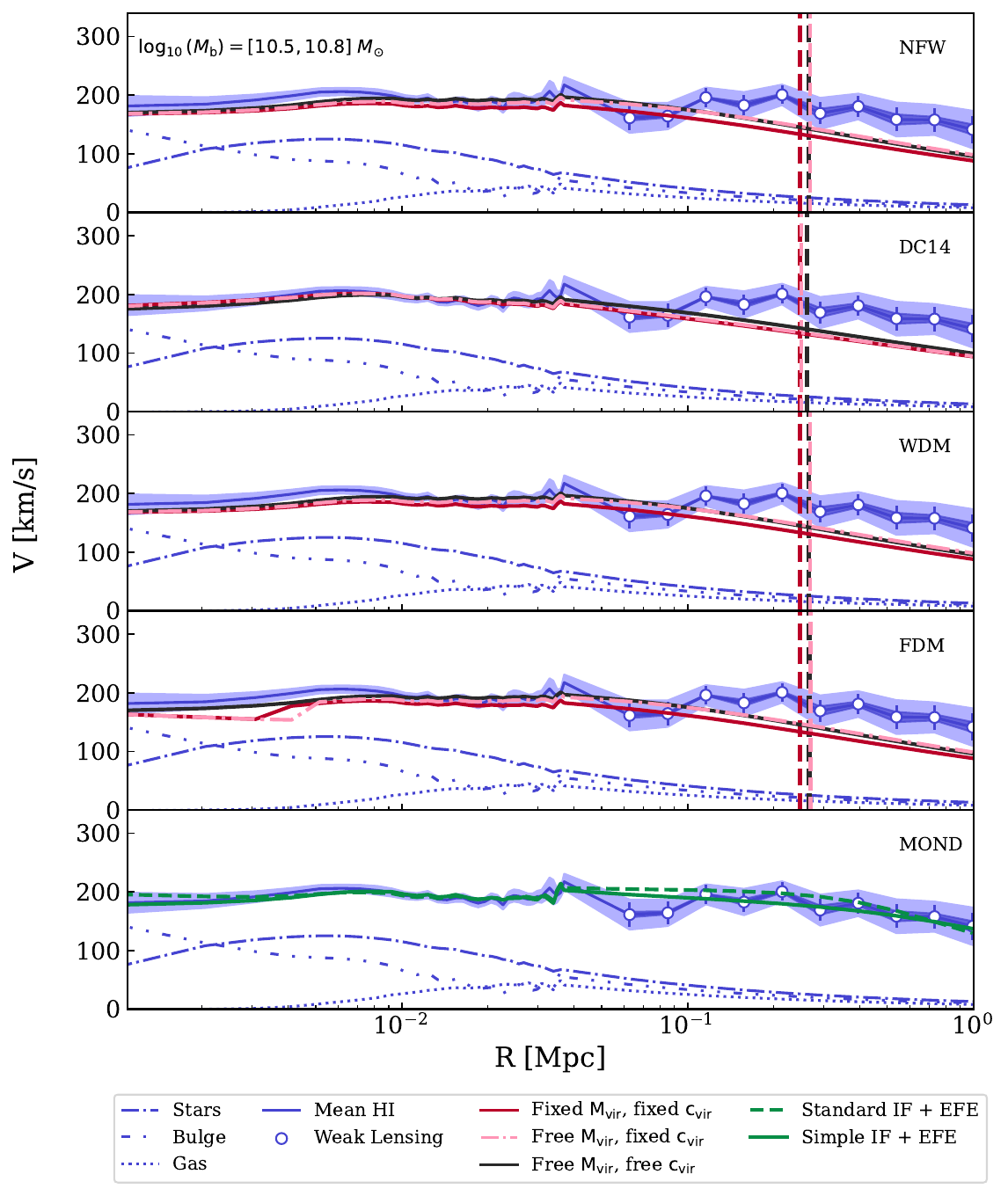}\hfill
  \caption{Same format as Figure 10 for mass bin $\log_{10}(M_{\mathrm{b}}/\mathrm{M_{\odot}}) = [10.5,10.8]$.}
  \label{fig:Mean_RC_Best_Fit_2}
\end{figure*}

\begin{figure*} 
  \centering
  \includegraphics[width=0.95\linewidth]{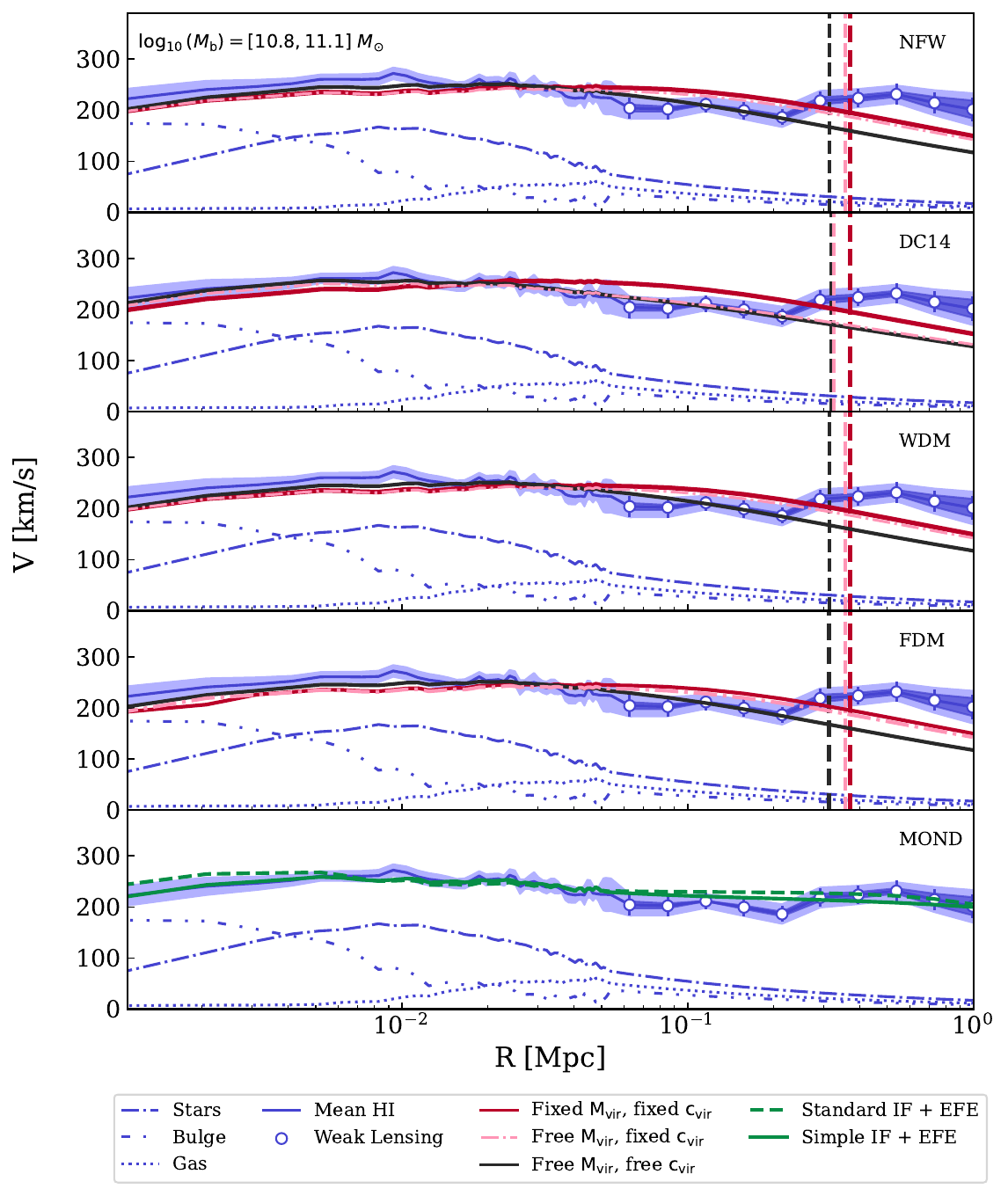}\hfill
  \caption{Same format as Figure 10 for mass bin $\log_{10}(M_{\mathrm{b}}/\mathrm{M_{\odot}}) = [10.8,11.1]$.}
  \label{fig:Mean_RC_Best_Fit_3}
\end{figure*}

\begin{figure*} 
  \centering
  \includegraphics[width=0.95\linewidth]{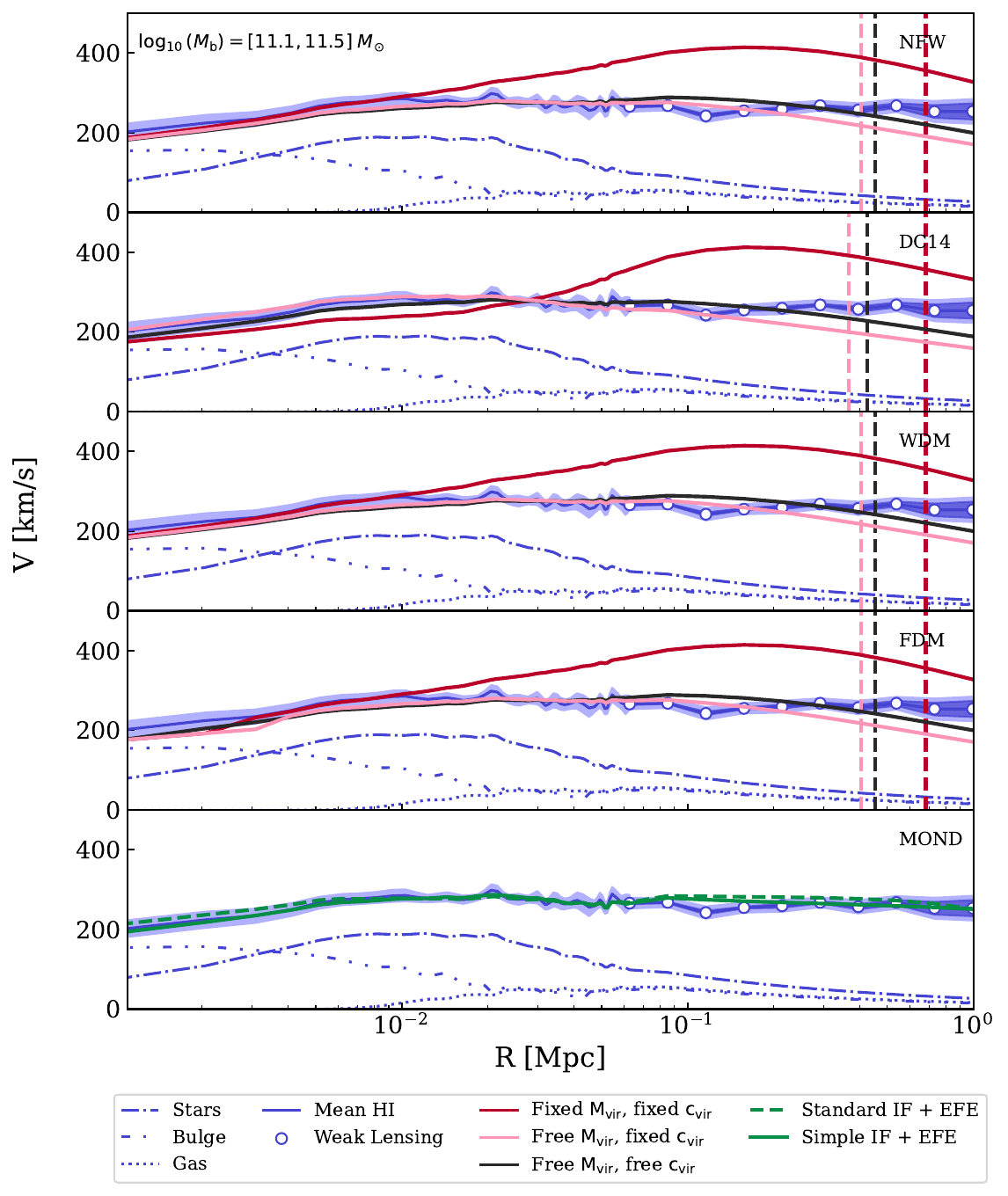}\hfill
  \caption{Same format as Figure 10 for mass bin $\log_{10}(M_{\mathrm{b}}/\mathrm{M_{\odot}}) = [11.1,11.5]$.}
  \label{fig:Mean_RC_Best_Fit_4}
\end{figure*}

Next, we take the mean velocity of all galaxies within this mass bin up until the largest measured radii of the group, and attach the weak lensing data from \citet{2024ApJ...969L...3M} to create a mean total rotation curve out to $1$ Mpc, as seen in Figures~\ref{fig:Mean_RC_Best_Fit_1}-\ref{fig:Mean_RC_Best_Fit_4}. We only include radial bins where there are three or more galaxies to avoid any artifacts from low sample size biases. We also exclude the inner two data points of the weak lensing data as the assumption of spherical symmetry in deprojecting the excess mass surface density is unreliable \citep{2024ApJ...969L...3M}. \citet{2024ApJ...969L...3M} also state that the measurements beyond $1$ Mpc do not have a reliable isolation criterion, so we only use the data up to this scale. The errors on the mean of the HI data RCs were calculated by $\sigma_{\mathrm{EOM}}=\sigma/\sqrt{N}$, where $\sigma$ is the standard deviation of the velocity measurements within each radial bin, and $N$ is the number of galaxies in that bin. In the MCMC process we account for any correlated uncertainties via the use of a covariance matrix \citep[][private communication]{2024ApJ...969L...3M}. The diagonal entries of these matrices are the squared stacked statistical uncertainties on the mean circular velocities, and are shown by the error bars in Figures~\ref{fig:Mean_RC_Best_Fit_1}-\ref{fig:Mean_RC_Best_Fit_4}. To calculate the HI covariance matrix we set the diagonal entries as the squared errors on the mean, $\sigma_{EOM}^{2}$, and the off-diagonal errors to zero. The total covariance matrix was then formed as a block-diagonal combination of the HI and weak-lensing covariance matrices, which was used in the MCMC likelihood evaluation (See Equation~\ref{eq:Cov}).

In Figures~\ref{fig:Mean_RC_Best_Fit_1}-\ref{fig:Mean_RC_Best_Fit_4} we fit the total mean rotation curves with the following models: NFW, DC14, WDM ($m_{\chi}=3\,\mathrm{keV}$), FDM ($m_{\chi}=10^{-22}\,\mathrm{eV}$), and MOND + EFE, with  $a_{0}=1.2\times 10^{-10} \:\mathrm{m\;s}^{-2}$ (assuming both the Simple and Standard IFs). In the MOND case we expect the EFE to be minimal due to the isolation criterion of the weak lensing data. The dark matter best-fit curves (red, pink, and black curves) are shown in the top four panels, and the MOND best-fit curves (green curves) are shown in the bottom panels of each figure. In the DM case we chose three halo parameter prescriptions, namely: i) Allowing both halo parameters to vary freely between the limits given in Table~\ref{table:free params} (black solid lines); ii) allowing $M_{\mathrm{vir}}$ to vary and fixing $c_{\mathrm{vir}}$ via the $c_{\mathrm{vir}}$–$M_{\mathrm{vir}}$ relation (pink solid and dot-dashed lines); and iii) taking the mean luminosity of the \texttt{SPARC} subsample to fix $M_{\mathrm{vir}}$ via the LHMR, then using the $c_{\mathrm{vir}}$–$M_{\mathrm{vir}}$ relation to fix $c_{\mathrm{vir}}$ (red solid curves). We do not include any scatter around the scaling relations as the error on the mean predicted by $\Lambda$CDM are expected to be extremely small, and the mean RC parameters are expected to lie on the mean scaling relation predictions.

We compare the three prescriptions of the best fit curves retrieved by fitting the full 1 Mpc range for each dark matter model and MOND, and present the reduced chi-squared statistics of each one in Table~\ref{tab:MeanRC_chi}. We immediately note the difficulty in simultaneously fitting the HI and weak lensing data with dark matter. Indeed, all the dark matter models, including the DC14 one, struggle to fully reproduce the larger portions of the mean RC. When anchored on the \texttt{SPARC} mean HI data below $\sim5$-$10$ kpc, the dark matter models are able to broadly align with the weak lensing data out to $\sim70-200\,\mathrm{kpc}$ depending on the baryonic mass bin, but then inevitably fall below the weak lensing data at larger scales, and cannot reproduce the weak-lensing data out to the virial radius (vertical lines in the tops panels of Figures~\ref{fig:Mean_RC_Best_Fit_1}-\ref{fig:Mean_RC_Best_Fit_4}). This is particularly evident for the cosmological models that prescribe the \citet{Dutton2014} $c_\mathrm{vir}-M_\mathrm{vir}$ prior (red and pink curves). This qualitative behaviour is more quantitatively reflected in the $\chi^{2}_{\mathrm{\nu}}$ values. Table~\ref{tab:MeanRC_chi} presents the reduced chi-squared statistics for each model and baryonic mass bin, computed for both the full mean rotation curves (left columns) and the weak-lensing data alone (right columns). While all the dark matter models produce acceptable total $\chi^{2}_{\mathrm{\nu}}$ when considering the total RCs, a different picture emerges when only the weak-lensing data are considered. When excluding the HI data and focusing only on the weak lensing data, the $\chi^{2}_{\mathrm{\nu},out}$ values show a tendency for dark matter to underperform at these scales. Although most $\chi^{2}_{\nu,\mathrm{tot}}$ values are close to unity, these statistics are largely driven by the HI data, due to the smaller number of weak lensing data points.

\begin{figure*} 
  \centering
  \includegraphics[width=\linewidth]{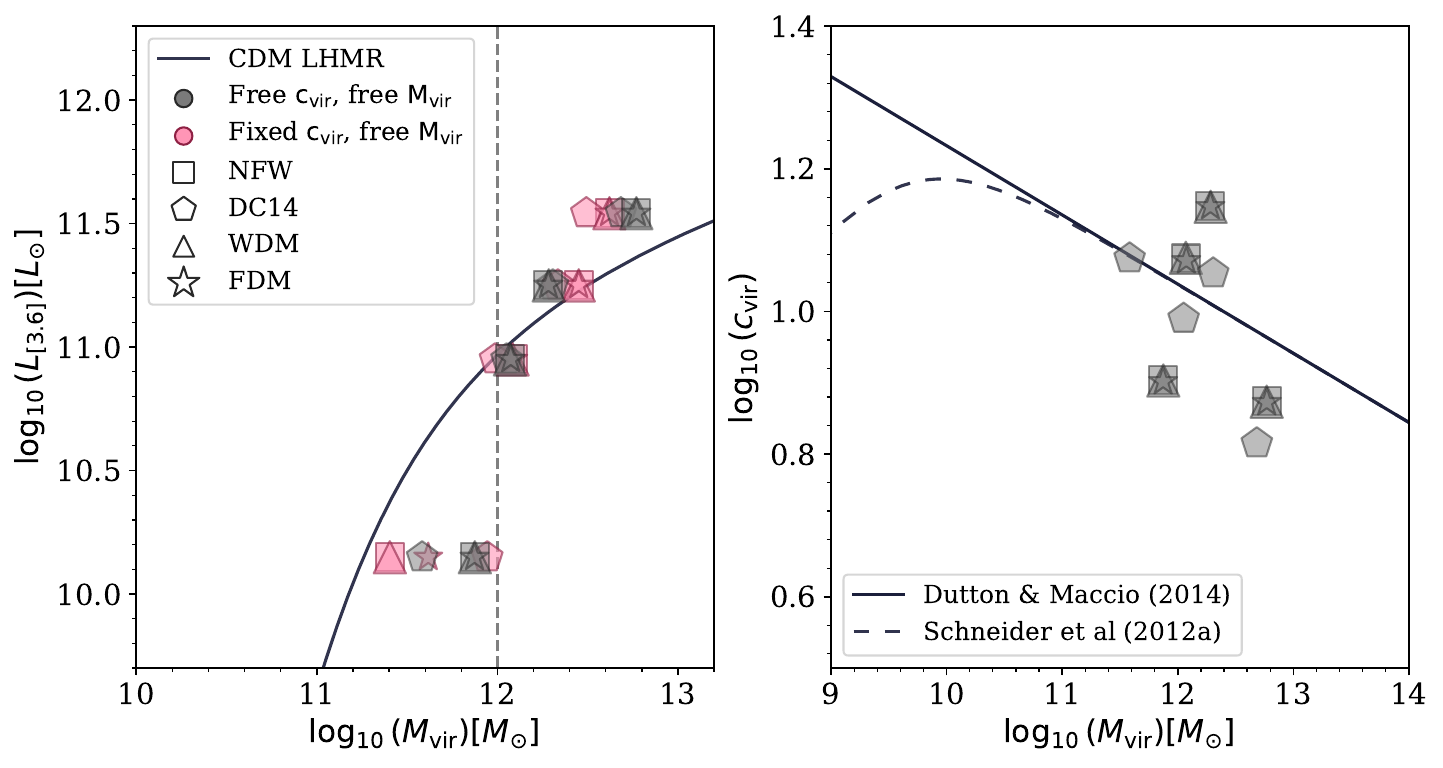}
  \caption{The best-fit halo parameters of each dark matter model against the LHMR (left panel) and the $c_{\mathrm{vir}}-M_{\mathrm{vir}}$ (right panel). The grey shapes show the halo parameters of the best-fit RCs (assuming $c_{\mathrm{vir}}$ and $M_{\mathrm{vir}}$ are free parameters) to the HI and weak lensing mean RCs out to $1$ Mpc. The pink shapes show the halo parameters corresponding to the RC fits when $M_{\mathrm{vir}}$ is free, and $c_{\mathrm{vir}}$ is fixed by the \citet{Dutton2014} $c_{\mathrm{vir}}-M_{\mathrm{vir}}$ relation. \textit{Left Panel:} best-fit halo parameters against the CDM (black line) LHMR. The vertical dashed line indicates the mass beyond which the DC14 model is extrapolated. \textit{Right Panel}: The best-fit halo parameters against the CDM \citet{Dutton2014} (black solid line) and the WDM \citet{10.1111/j.1365-2966.2012.21252.x} (black dashed line) $c_{\mathrm{vir}}-M_{\mathrm{vir}}$ relation.}
  \label{fig:Mean_RC_Best_Fit_Scaling_Relations}
\end{figure*}

In the left-hand panel of Figure~\ref{fig:Mean_RC_Best_Fit_Scaling_Relations}, we compare the scaling relation predictions with the best-fitting $M_{\mathrm{vir}}$ values as a function of the mean luminosity of each mass bin. Over the halo mass range considered here, the theoretical predictions for CDM, WDM, and FDM are effectively identical, since the halo mass functions are not significantly suppressed for $M_{\mathrm{vir}}>10^{10.5} \:\mathrm{M_{\odot}}$. The best-fitting halo masses lie systematically above the expected relation in mass bins 1 and 2, while in mass bin 3 and 4 they fall below the LHMR prediction owing to the turnover in the halo abundance matching relation. Such a discrepancy may be at least partly explained by the fact that late-type galaxies allegedly follow steeper scaling relations with halo mass than the total population \citep[e.g.,][]{2017MNRAS.470..651R, 2019A&A...629A..59P, 2025A&A...704A.244F}. In the right panel of Figure~\ref{fig:Mean_RC_Best_Fit_Scaling_Relations} we plot the best-fit $M_{\mathrm{vir}}$ and $c_{\mathrm{vir}}$ values of each model. We also show the WDM $c_{\mathrm{vir}}$-$M_{\mathrm{vir}}$ relation from \citet{10.1111/j.1365-2966.2012.21252.x}. The grey shaped markers represent the best-fit $M_{\mathrm{vir}}$ and $c_{\mathrm{vir}}$ values respectively of each dark matter model in each baryonic mass bin when both halo parameters are allowed to be free. All DM models tend to favour best-fit concentrations significantly above or below the mean $c_{\mathrm{vir}}$-$M_{\mathrm{vir}}$ relation predicted by CDM. The DC14 tends to predict concentrations closer to the mean relation but still outside of the expected error on the mean, apart from bin 1 which can reproduce the mean relation. It is immediately clear that none of the dark matter models can faithfully reproduce the flatness of the weak lensing data, and simultaneously lie in agreement with the mean scaling relations. For example, in bin 1 the NFW, WDM, and FDM models require more massive, and less concentrated haloes than predicted by the scaling relations (see grey symbols in Figure~\ref{fig:Mean_RC_Best_Fit_Scaling_Relations}). It must be noted that the \citet{Dutton2014} $c_{\mathrm{vir}}$-$M_{\mathrm{vir}}$ relation used in this work is from DM-only simulations, whereas the $c_{\mathrm{vir}}$-$M_{\mathrm{vir}}$ relation predicted by the hydrodynamical simulation of TNG100 is steeper than that of \citet{Dutton2014}. Although feedback is expected the change the $c_{\mathrm{vir}}$-$M_{\mathrm{vir}}$ relation, the fits of the dark matter models to each baryonic bin mean RC do not follow the expectation of either relations, due to their large scatter. Since the radial scales of the weak lensing data are in the dark matter dominated regime, the normalisation (velocity) of the weak lensing data strongly constrains the halo mass. The inferred halo masses remain nearly identical across all models (with DC14 preferring slightly smaller halo masses in bin 1) within any given baryonic mass bin, as reflected by the parameter values reported in Table~\ref{tab:MeanRC_params}. The concentration is then constrained by the slope of the HI data and the flatness of the weak lensing data. Although in bins 1 and 4 lower concentrations are favoured to reproduce a flatter outer RC, this is not required in bins 2 and 3 (see the right panel of Figure~\ref{fig:Mean_RC_Best_Fit_Scaling_Relations}), where in fact larger concentrations are preferred. This is most likely due to bins 2 and 3 having flatter HI curves than in bins 1 and 4. When the concentrations are fixed to the mean scaling relation of \citet{Dutton2014}, the dark matter rotation curves tend to predict a stronger downturn than in the weak lensing data in bins 1 and 4, whereas they are similar in bins 2 and 3. In bin 1 when $c_{\mathrm{vir}}$ is fixed, the DC14 is able to reproduce the weak lensing data ($\chi^{2}_{\nu,out}=0.691$) only because the fit to the HI region has been sacrificed, resulting in $\chi^{2}_{\nu,tot}=4.06$ for the full RC. This further highlights the tension between HI and weak lensing data fits for the dark matter models of this work. In bin 2 the FDM model presents a break feature in the best-fit RC when the concentration is fixed. As detailed in Section~\ref{sec:FDM}, this arises from the continuity condition failing to be satisfied due to constraints imposed on the parameter space. This feature drives to poor fit, resulting in a reduced chi-squared value of $\chi^{2}_{\nu,tot}\sim2.8-3.5$.

We also fit the RCs using the 200 times critical density definition (200c) of halo mass and concentration, to test whether differing overdensity choices affect the conclusions of Figure~\ref{fig:Mean_RC_Best_Fit_Scaling_Relations}. We note that using the 200c overdensity definition does not change the overall argument. The discrepancy between NFW, WDM, and FDM models against the \citet{Dutton2014} $c_{\mathrm{vir}}-M_{\mathrm{vir}}$ relation is slightly reduced, however the rotation curve fits still lie outside of the mean $c_{\mathrm{vir}}$-$M_{\mathrm{vir}}$ relation. In summary, all of the dark matter models may only be able to reproduce the mean RCs up to scales of $\sim100\,\mathrm{kpc}$, thus struggling to fully reproduce the flatness of the weak lensing data out to 1 Mpc.

\renewcommand{\arraystretch}{1.25}
\setlength{\tabcolsep}{11pt}
\begin{table*}
\begin{tabular}{llcccccccc}
\hline
\multicolumn{1}{|l|}{\textbf{Model}} & \multicolumn{1}{l|}{\textbf{Parameters}} & \multicolumn{4}{c|}{\textbf{$\chi^{2}_{\mathrm{\nu, tot}}$}}& \multicolumn{4}{c|}{\textbf{$\chi^{2}_{\mathrm{\nu},out}$}} \\ 
&& \textbf{Bin 1} & \textbf{Bin 2} & \textbf{Bin 3} & \textbf{Bin 4} & \textbf{Bin 1} & \textbf{Bin 2} & \textbf{Bin 3} & \textbf{Bin 4}\\ \hline
NFW  & Fixed $M_{\mathrm{vir}}$, fixed $c_{\mathrm{vir}}$ & 5.55& 3.20 & 2.35 & 31.49 & 4.56 & 1.61 & 18.7 & 140\\
  &  Free $M_{\mathrm{vir}}$, fixed $c_{\mathrm{vir}}$ &  3.01 & 1.83 & 2.35 & 1.11 & 3.75 & 2.85 & 18.1 & 13.0\\
  &  Free $M_{\mathrm{vir}}$, free $c_{\mathrm{vir}}$ &  0.632 & 1.48 & 1.83 & 1.16 & 0.756 & 3.22 & 14.2 & 25.3\\
DC14 &  Fixed $M_{\mathrm{vir}}$, fixed $c_{\mathrm{vir}}$ & 4.81 & 1.44 & 2.66 & 18.3 & 4.45 & 1.55 & 19.6 & 131\\
  &  Free $M_{\mathrm{vir}}$, fixed $c_{\mathrm{vir}}$ & 4.06 & 1.37 & 1.26 & 1.69 & 0.691 & 1.82 & 12.8 & 5.60\\
  &  Free $M_{\mathrm{vir}}$, free $c_{\mathrm{vir}}$ & 1.41 & 1.18 & 1.17 & 0.77 & 2.39 & 2.73 & 13.6 & 17.0\\
WDM &  Fixed $M_{\mathrm{vir}}$, fixed $c_{\mathrm{vir}}$ & 5.63 & 3.21 & 2.35 & 31.5 & 4.56 & 1.61 & 18.7 & 140\\
  &  Free $M_{\mathrm{vir}}$, fixed $c_{\mathrm{vir}}$ & 3.07 & 1.83 & 1.36 & 1.11 & 3.77 & 2.85 & 18.1 & 13.0\\
  &  Free $M_{\mathrm{vir}}$, free $c_{\mathrm{vir}}$ & 0.628 & 1.48 & 1.83 & 1.16 & 0.752 & 3.21 & 14.3 & 25.4\\
FDM &  Fixed $M_{\mathrm{vir}}$, fixed $c_{\mathrm{vir}}$ & 5.61 & 3.53 & 2.40 & 31.5 & 5.61 & 1.61 & 18.7 & 140\\
  &  Free $M_{\mathrm{vir}}$, fixed $c_{\mathrm{vir}}$ & 1.30 & 2.84 & 2.22 & 1.20 &  1.30 & 3.25 & 21.1 & 14.6\\
  &  Free $M_{\mathrm{vir}}$, free $c_{\mathrm{vir}}$ & 0.650 & 1.51 & 1.86 & 1.19&  0.650 & 4.90 & 4.75 & 29.0\\
\hline

\multirow{2}{2.1cm}{Standard IF + EFE}&Free $\Upsilon_{*}$, free $a_{\mathrm{e}}$,&&&&&&&& \\ 
& fixed $a_{0}$  & 0.193 & 0.564 & 0.787 & 0.410 & 0.364 & 0.981 & 1.725 & 2.264\\
\multirow{2}{2.1cm}{Simple IF + EFE}&Free $\Upsilon_{*}$, free $a_{\mathrm{e}}$,&&&&&&&& \\ 
& fixed $a_{0}$  & 0.245 & 0.538 & 0.451 & 0.264 & 0.707 & 0.671 & 1.124 & 0.954\\
\hline                                        
\end{tabular}
\caption{The reduced chi-square, $\chi^{2}_{\mathrm{\nu}}$, statistic of the dark matter and MOND fits to the mean RCs of baryonic mass bins 1-4, defined by $\log_{10}(M_{\mathrm{b}}/\mathrm{M_{\odot}})=[9.0,10.5],\;[10.5,10.8],\;[10.8,11.1],\;[11.1,11.5]$ respectively. The $\chi^{2}_{\mathrm{\nu}}$ values are calculated for the full RCs, $\chi^{2}_{\mathrm{\nu},tot}$ (left columns), and the outer weak-lensing data, $\chi^{2}_{\mathrm{\nu},out}$ (right columns) from \citet{2024ApJ...969L...3M}.}
    \label{tab:MeanRC_chi}
\end{table*}

For MOND, we fix the acceleration constant at $a_{0}=1.2\times 10^{-10} \:\mathrm{m\;s}^{-2}$, and fit for the mean mass-to-light ratio, $\Upsilon_{*}$, of each baryonic mass bin, which is given a flat prior between $0.4-1.0\,\mathrm{M_{\odot}/L_{\odot}}$. The best-fit of the Standard and Simple IFs are given by the dashed and solid green lines respectively, where the free parameters $\Upsilon_{*}$, and $a_{\mathrm{e}}$ are given in Table~\ref{tab:MeanRC_params}. For each bin $a_{\mathrm{e}}$ results to be extremely small, capped at a value of $a_{\mathrm{e}}\sim2.8\%$ of the characteristic MOND acceleration, $a_{0}$. Such a tiny EFE contribution is expected in isolated galaxies with very flat RCs. We use a flat prior on $\Upsilon_{*}$ instead of fixing the same value across all mass bins since the late-type spirals and dwarfs tend to have a bluer colour and therefore lower $\Upsilon_{*}$ than bulge-dominated galaxies with redder colours. We see that the Standard IF generally predicts larger $\Upsilon_{*}$ values than the Simple IF, this is in agreement with the trend of individual galaxy fits of \texttt{SPARC} galaxies, however both produce good fits to the data with values that fall within the range expected from stellar population synthesis models \citep{2022AJ....163..154S,2025AJ....169..186D}. Our best-fit MOND models are able to fully reproduce the mean RC of \citet{2024ApJ...969L...3M} in all mass bins out to 1 Mpc. We generally find that the IFs produce degenerate results at all scales, and systematically fit the mean HI and weak-lensing data very well with $\chi^{2}_{\mathrm{\nu},tot}\sim0.2-0.8$. However, if a discrepancy between MOND and the mean HI data were present, this would not necessarily be indicative of a shortcoming of MOND. Since the mean HI data are calculated from a variety of morphologies, stellar mass distributions, and mass-to-light ratios, the resulting mean baryonic distribution does not represent a self-consistent galaxy, but rather a composite of different systems. The Standard IF + EFE model has shown to fit the individual rotation curves very well, and any small residuals (which are not present here) around the mean HI rotation curves could be a result of this averaging effect in the mean baryonic components rather than a failure of MOND itself.

\begin{table*}
    \centering
    \small
    \setlength{\tabcolsep}{3pt}
    \begin{tabular}{|c|cc|cc|cc|cc|}
        \hline

        & \multicolumn{2}{|c|}{Bin 1} 
        & \multicolumn{2}{c|}{Bin 2} 
        & \multicolumn{2}{c|}{Bin 3} 
        & \multicolumn{2}{c|}{Bin 4} \\
        \hline
        
        & $\log_{10}(M_{\mathrm{vir}}/M_{\odot})$ & $c_{\mathrm{vir}}$
        & $\log_{10}(M_{\mathrm{vir}}/M_{\odot})$ & $c_{\mathrm{vir}}$
        & $\log_{10}(M_{\mathrm{vir}}/M_{\odot})$ & $c_{\mathrm{vir}}$
        & $\log_{10}(M_{\mathrm{vir}}/M_{\odot})$ & $c_{\mathrm{vir}}$ \\
        \hline 
        NFW & $11.87$ & $8.02$ & $12.07$ & $11.86$ & $12.28$ & $14.06$ & $12.77$ & $7.48$\\
        DC14 & $11.58$ & $11.87$ & $12.05$ & $9.77$ & $12.31$ & $11.31$ & $12.68$ & $6.53$\\
        WDM & $11.87$ & $8.00$ & $12.07$ & $11.89$ & $12.28$ & $14.02$ & $12.77$ & $7.47$\\
        FDM & $11.87$ & $7.97$ & $12.07$ & $11.78$ & $12.28$ & $14.03$ & $12.77$ & $7.45$\\
        \hline

        
        & $\Upsilon_{*}$ & $a_{\mathrm{e}}$
        & $\Upsilon_{*}$ & $a_{\mathrm{e}}$
        & $\Upsilon_{*}$ & $a_{\mathrm{e}}$
        & $\Upsilon_{*}$ & $a_{\mathrm{e}}$ \\
        \hline 
        Standard IF + EFE, Fixed $a_{0}$ & $0.674$ & $0.00407$& $0.740$ &  $0.0335$& $0.827$ & $0.0164$ & $0.760$ & $0.0254$ \\
        Simple IF + EFE, Fixed $a_{0}$ & $0.494$ & $0.000917$& $0.548$ & $0.00884$& $0.622$ & $0.00260$& $0.563$ & $0.00282$\\
        \hline
    \end{tabular}
    \caption{The maximum-likelihood parameters of the best-fit rotation curves (black curves in Figures~\ref{fig:Mean_RC_Best_Fit_1}-\ref{fig:Mean_RC_Best_Fit_4})) of each dark matter model and MOND (Standard/Simple IF + EFE with $a_{0}=1.2\times 10^{-10} \:\mathrm{m\;s}^{-2}$), obtained from the mean rotation curve fits in four baryonic mass bins defined by $\log_{10}(M_{\mathrm{b}})=[9.0,10.5],\;[10.5,10.8],\;[10.8,11.1],\;[11.1,11.5]\;M_{\odot}$ in ascending order. Note that $\Upsilon_{*}$ and $a_{\mathrm{e}}$ are in units of $\mathrm{M_{\odot}/L_{\odot}}$ and $10^{-10}\;\mathrm{ms}^{-2}$ respectively.}
    \label{tab:MeanRC_params}
\end{table*}

\section{Discussion} \label{sec:Discussion}

In this work we revisited the performance of different DM and MOND models against the RCs of \texttt{SPARC} and also the most recent ones up to 1 Mpc derived from weak lensing data. We also presented the scaling relation between the $3.6\;\mathrm{\mu m}$-band absolute luminosity and the virial halo mass of galaxies, and used the inferred halo masses of dark matter and MOND models to discuss the predictive power of this relation. In this section we discuss our results, compare with other findings in the literature, and highlight the impact of possible systematics and assumptions.

\subsection{NFW vs DC14} 

Here we have analysed a variety of popular dark matter models. In particular, we adopted the NFW profile, which is a fit to N-body simulations, and the DC14 model, which is a fit to hydrodynamic simulations inclusive of baryonic feedback. The latter is naturally more flexible, allowing for the generation of cored profiles when the stellar-to-halo mass ratio is higher than $0.01\%$. Despite the added flexibility, the DC14 model is only marginally more successful than the NFW profile, with a $\chi_{\nu}^{2}$ lower by just $\sim0.45$ and a $\Delta\mathrm{BIC}>6$ for just $33.1\%$ of the sample when fixing distance and inclination (or $\sim0.33$ and $31.4\%$ when observational uncertainties in distance and inclination are accounted for). The DC14 is particularly strong in reproducing a small scatter around the luminosity-halo mass relation, as also noted by \citet{2017MNRAS.466.1648K} and \citet{2017A&A...605A..55A}, although, as with any other model, it tends to predict systematically larger halo masses at fixed luminosity than what is predicted by the LHMR. \citet{2026arXiv260107799B} recently argued that \texttt{SPARC} galaxies may be biased towards lower circular velocities at fixed halo mass, which in turn could induce larger halo masses at fixed galaxy stellar mass (their Figure 6). Such a trend may help in reconciling the offset we observe in the LHMR relation (Figure~\ref{fig:LHMR}), although there is no independent confirmation of such a bias in \texttt{SPARC} galaxies, as they appear consistent with established scaling relations \citep[e.g.,][]{2016ApJ...816L..14L}. We will further investigate this point in future work.

We also explored whether the mean HI curves are biased by rotation curve morphology. Rising RCs that have not been measured beyond the radius of the peak velocity may overestimate halo masses and incorrectly infer other halo parameters due to degeneracies. The velocities of the sloped subsample (defined by the velocity gradient $\gamma>0.1$) are systematically larger than that of the flat subsample, although only moderately. We find that the best-fit values of the mean halo masses competing to each galaxy subsample differs by just $\sim0.1$ dex for each dark matter model, which is insufficient to resolve the tension between the scaling relations and the RCs data.

\subsection{Warm Dark Matter} \label{sec:WDM_dis}

In this work we also tested the performance of the Warm Dark Matter model assuming a particle mass of $m_{x}=3$ keV and $m_{x}=1$ keV. When $m_{x}\geq3$ keV the halo mass function is not dampened significantly at the mass scales considered here. The stronger suppression of small-scale structures in the $m_{x}=1$ keV case results in a stronger reduction of the inner velocities \citep{2012MNRAS.424.1105M}. We found that WDM models yield very similar results to a classical NFW profile in terms of rotation curve fitting and scaling relations (see Table~\ref{table:BIC} and Figure~\ref{fig:LHMR}). The WDM model (with $m_{x}=3$ keV) similarly to the NFW profile, cannot produce a rotation curve that is consistent with the mean one up to 1 Mpc, with an implied mean host halo mass close to the one predicted by abundance matching (Figure~\ref{fig:Outliers}). We note that although the 1 keV WDM is somewhat in tension with the current constraints from the Lyman alpha forest \citep{2023PhRvD.108b3502V}, it is still too early to exclude such models with confidence given the sizeable model-dependent assumptions characterising the observational constraints (e.g., simulation design, thermal history, environmental conditions).

\subsection{Fuzzy Dark Matter} \label{sec:FDMDis}

We included in our analysis a FDM model with a fixed particle mass of $m_{\chi}=10^{-22}$ eV. As the mass density is only modulated by the core radius, $r_{c}$, and transition radius, $r_{t}$, (see Section~\ref{sec:FDM}), by fixing the mass, the minimisation process has less flexibility and can sometime induce a discontinuity around the transition radius. Indeed, \citet{2023MNRAS.523.3393K} assumed a free particle mass within the interval $\log_{10}(m_{\chi})=[-3,3]\times10^{-22}$ eV concluding that across the \texttt{SPARC} sample not one singular $m_{\chi}$ value could explain all of the rotation curve shapes at a $95\%$ confidence level. Although \citet{2023MNRAS.523.3393K} find that the NFW model outperforms the FDM model, we expect this is likely due to the added flexibility in their RC fits when allowing the particle mass to vary for each galaxy. FDM models have also been disfavoured on the basis of their central mass density values and trends of core radius with halo mass in tension with current observational constraints \citep{2020ApJ...904..161B}.

\subsection{MOND vs Dark Matter} \label{Comparison}

In this work we found that MOND is able to reproduce a substantial fraction of the rotation curves in our sample of 147 late-type spiral galaxies, despite employing fewer free parameters than the dark-matter-based models. Owing to this reduced flexibility, the median reduced chi-square values, $\chi^2_{\mathrm{\nu}}$, for both the Simple and Standard interpolation functions are, at face value, systematically higher than those obtained for any of the dark matter models. However, we emphasise that such a direct comparison is not statistically well defined, since the $\chi^2_{\mathrm{\nu}}$ statistic does not account for the fundamentally different roles played by the free parameters in MOND and dark matter frameworks. In particular, MOND is constrained by a small set of global, physically motivated constants, whereas dark matter halo models rely on multiple object-to-object adjustable parameters. This distinction is analogous to the argument made by \citet{2021A&A...646L..13L} regarding the gravitational constant, where parameters with different physical status and prior constraints should not be treated on an equal statistical footing within model comparison metrics. In our sample, the median difference in $\chi^2_{\mathrm{\nu}}$ between the best performing MOND configuration and the worst performing dark matter model is only $\sim0.4$ which can be easily accounted for via residual observational statistical and systematic uncertainties.

\subsection{Extended Rotation Curves: MOND or Dark Matter?} \label{sec:Ext}

Weak lensing data have been used to constrain dark matter halo density profiles out to 1 Mpc. \citet{2024ApJ...969L...3M} argued that there is tension between the expectations from the $\mathrm{\Lambda}$CDM NFW density profile and the observed weak lensing velocity curves at these large distances. We note that \citet{2024ApJ...969L...3M} used a spherical overdensity of $\Delta_{c}=200$ which is different to the value of $\Delta_{c}\approx100$ \citep{1998ApJ...495...80B} used in this work. By using $\Delta_{c}\approx100$, as in this work, the halo masses are larger than predicted when using $\Delta_{c}\approx200$, but the shape of the rotation curve remains similar. This may explain why the virial radii of our best fit haloes (as seen in Figures~\ref{fig:Mean_RC_Best_Fit_1}-\ref{fig:Mean_RC_Best_Fit_4}) are larger than those reported in \citet{2024ApJ...969L...3M}.

As seen in Figures~\ref{fig:Mean_RC_Best_Fit_1}-\ref{fig:Mean_RC_Best_Fit_4}, the MOND curves naturally produce flat RCs out to large scales. On the other hand, all of the dark matter models may be able to reproduce the HI data but are unable to reproduce the weak lensing measurements up to the virial radius as they tend to inevitably fall short at larger scales and with best-fit values of concentration often lying outside of the expected range of mean values (Figure~\ref{fig:Mean_RC_Best_Fit_Scaling_Relations}).

Since \texttt{SPARC} is dominated by late-type galaxies (LTGs), we also test whether selecting the LTG-only weak lensing rotation curves has any effect on the fits of dark matter models. We independently verified that the best-fit RCs to the \texttt{SPARC} and LTG-only weak lensing data remains largely unchanged from that of the LTG+ETG weak lensing fits in Figures~\ref{fig:Mean_RC_Best_Fit_Scaling_Relations}. The tension between the outskirts of weak lensing data and dark matter models remains, thus leaving our conclusion unchanged.

All in all, our analysis shows that MOND can reproduce current weak lensing data up to 1 Mpc, whereas dark matter models struggle to generate continuously flat rotation curves beyond the virial radius whilst still being consistent with the mean expected values of concentration and halo mass. In a companion paper \citep[][submitted]{Jeong2026}, we will further explore the detailed performance of a $\Lambda$CDM model with baryonic feedback using the direct outputs of a state-of-the-art hydrodynamic simulation to compare with weak lensing data. In this work, we will assess the extent to which unbound material in the vicinity of a halo contributes to the measured rotation curve at scales approaching 1 Mpc, and whether such contributions can account for the apparent flatness of weak lensing rotation curves.

\subsection{Mass-to-light ratio} \label{ML}

To simplify the rotation curve fitting procedure the mass-to-light ratio, $\Upsilon_{*}$, has been assumed to be constant across the radial extent of each galaxy, although we discussed above that some models could benefit from a gradient in $\Upsilon_{*}$. A $\Upsilon_{*}$ varying with galactocentric distance has been explored in several papers \citep[e.g.,][]{2000ApJ...534..670T, 2010A&A...521A..82P, Lelli_2016, 2022MNRAS.510.5639M, 2024MNRAS.532..538K} with evidence supporting a $\Upsilon_{*}$ decreasing towards the outskirts of galaxies. \citet{2025MNRAS.536.2672L} for example analysed massive early-type galaxies from the \texttt{TNG100} simulation finding that these systems tend to have more centrally concentrated stellar mass profiles than their light distribution. Assuming a constant ratio $\Upsilon_{*}$, they find, would  generally overestimate the mass-to-light ratio whilst underestimate the central dark matter fractions. On the other hand, \citet{2024MNRAS.532..538K} argued that the $\Upsilon_{*}$ of low mass galaxies in the \texttt{FIRE} simulation increases with distance from the galactic centre, suggesting that stellar rotation curves may be underestimated at large radii. We will discuss these possibilities in \citet[][submitted]{Jeong2026}. In the presented mean rotation curves of Section~\ref{sec:MeanRC} we find that fitting for the mean mass-to-light ratio and fixing the canonical acceleration constant value, results in good fits to the weak-lensing data at all scales, and if discrepancies were present between the MOND rotation curves and the mean HI data, they could be explained by the effect of averaging the baryonic component of many different systems.

\section{Conclusions}

The large-scale behaviour of galaxy rotation curves requires to either invoke the presence of additional mass beyond the baryonic mass of galaxies, or to modify the gravity laws at low accelerations. In this work we have addressed this key issue by analysing the RCs of $147$ late-type spirals from the \texttt{SPARC} sample using the Navarro-Frenk-White (NFW) profile, the empirical baryonic feedback profile (DC14) from \citep{Cintio_2014}, Warm Dark Matter (WDM), Fuzzy Dark Matter (FDM), and Modified Inertia formulations of Modified Newtonian Dynamics (MOND) models, for the latter adopting the standard and simple interpolation functions. We used an MCMC approach to infer the best-fit values of the parameters of these models. We compared these parameters to the predictions of the Luminosity-Halo-Mass Relation (LHMR) derived via abundance matching between the local galaxy luminosity function at $3.6\mathrm{\mu m}$ \citep{2009ApJ...697..506D} and the halo mass function from \citet{2008ApJ...688..709T}. We then extended our analysis to the mean rotation curve of galaxies up to 1 Mpc through the inclusion of weak-lensing data from \citet{2024ApJ...969L...3M}. Our main findings can be summarised as follows:

\begin{itemize}
    \item The individual RCs of our subsample of $147$ late-type spirals are very well reproduced by the baryonic feedback model, DC14, from \citet{Cintio_2014} (see Table~\ref{table:Chi2vals}), with a median $\chi^{2}_{\mathrm{\nu}}=0.934$ when distance and inclination are fixed, which is lower than all other models. However, the difference between the median $\chi^{2}_{\mathrm{\nu}}$ of the other models is relatively small and cannot be used by itself to firmly exclude any of the models considered in this work. 
    \item The observational uncertainties effecting the baryonic velocity components, such as distance, inclination, and mass-to-light ratio hold more weight in the MOND scenario than for dark matter. It is important to account for these uncertainties when modelling the rotation curves of MOND.
    \item When deploying the BIC statistic to measure the relative performance of each model weighted against their number of free parameters, we still find that the data tend to prefer the DC14 model, although moderately, with only $\sim33\%$ of the sample strongly preferring DC14 over NFW or $\sim34\%-35\%$ over WDM (Table~\ref{table:BIC} and Figure~\ref{fig:BIC}). However, when only considering dwarf galaxies the strong preference for DC14 increases to $\sim40\%-50\%$ of the dwarf-only subsample.
    \item The Simple and Standard MOND interpolation functions yield values of $\chi^2_{\mathrm{\nu}}$ that are, at face value, higher than those obtained for all dark matter models when $a_0$ is treated as a free parameter (see Table~\ref{table:Chi2vals}). However, we note that a direct comparison between MOND and dark matter models is not strictly meaningful, as common statistical indicators implicitly weigh all parameters equally despite their differing physical significance and influence on the RC fits.
    \item Accounting for the uncertainties in distance and inclinations of each galaxy increases the $\chi^2_{\mathrm{\nu}}$ statistics of all dark matter models but decreases them for the MOND variations with less than 5 free parameters, bringing them only marginally higher (a factor of $\sim2$) than those from dark matter models (Table~\ref{table:Chi2vals}, right column).  There is no significant differences in the BIC statistic distributions of dark matter models with the addition of the observational uncertainties. 
    \item The addition of a $c_{\mathrm{vir}}$ prior around the \citet{Dutton2014} $c_{\mathrm{vir}}-M_{\mathrm{vir}}$ relation increases the median $\chi^{2}_{\mathrm{\nu}}$ (see Figure~\ref{fig:Chi2_DuttonPrior}) even though it reduces the scatter around the predicted scaling relations. This implies that dark matter models may require more extreme halo parameters to explain the diversity of observed RCs.
    \item We compare the maximum likelihood $M_{\mathrm{vir}}$ inferences against the predicted LHMR from abundance matching (see Figure~\ref{fig:LHMR}), and find a systematic preference for larger halo masses than predicted by these models. The DC14 model incorporating baryonic feedback tends to reduce the predicted scatter though it still shows a systematic preference for larger halo masses than what predicted at face value by the LHMR relation (Figures~\ref{fig:LHMR} and \ref{fig:Mean_RC_Best_Fit_Scaling_Relations}), which could also signal a shortcoming in the abundance matching procedure, a large contribution of galaxies with rising rotation curves, or possibly even a bias in the \texttt{SPARC} data. 
    \item When comparing with the mean rotation curves of the combined \texttt{SPARC} and weak lensing data of galaxies out to 1 Mpc, we find that MOND can naturally reproduce the flatness of the weak lensing data, whilst simultaneously fitting the mean HI velocities of all mass bins. Regardless of this, any tension between the mean HI data and MOND can perhaps be reconciled by considering the averaging effects over systems of different mass-to-light ratios in the \texttt{SPARC} sample.
    \item All the dark matter models explored here struggle to reproduce the mean rotation curves from HI and weak lensing data up to the virial radius and also decline too quickly beyond this scale (Figures~\ref{fig:Mean_RC_Best_Fit_1}-\ref{fig:Mean_RC_Best_Fit_4}). In addition, the best-fit values of concentration and/or halo mass frequently do not align with expectations from the mean large-scale scaling relations (Figure~\ref{fig:Mean_RC_Best_Fit_Scaling_Relations}).
    \end{itemize}

In summary, in this work we showed that current data from local individual galaxy rotation curves are unable to clearly pinpoint with high confidence one model over another. The inner regions of the rotation curves probed by the \texttt{SPARC} data may favour, although only moderately, baryonic feedback CDM models, like the one put forward by DC14. On the other hand, MOND is able to naturally reproduce the mean weak lensing galactic rotation curves up to 1 Mpc, which DM models struggle to do.

We expect that upcoming data from \texttt{BIG-SPARC} \citep{2024arXiv241113329H}, and from the Square Kilometre Array (SKA) will statistically strengthen the predictive power of the models \citep{2026arXiv260807290L} used in this work, whilst alleviating the possibility for observational biases in the calibration of mean rotation curves. However, as the weak-lensing data show greater discriminatory power, favouring MOND on scales beyond $100\,\mathrm{kpc}$, future weak-lensing surveys will be crucial for improving the robustness and precision of these constraints. In \citet{Jeong2026}, we further investigate in full hydrodynamic simulations the contribution of mass external to the bound halo to weak-lensing mass measurements, and its role in the observed flattening of the weak-lensing signal at scales of $\sim1\,\mathrm{Mpc}$, perhaps reconciling some of the tension between dark matter fits and the weak lensing data. Also in future work we will explore the performance of the dark matter and MOND models on other relevant galactic scaling relations to supplement the results of this paper.

\section*{Acknowledgements}

We acknowledge funding from the Leverhulme Trust (RPG-2022-57). HF acknowledges support at Fudan University from the Shanghai Super Post-doctoral Excellence Program grant No. 2024008. ADC is supported by the Agencia Estatal de Investigación, under the 2023 call "Ayudas
para Incentivar la Consolidación Investigadora", grant number CNS2023- 144669, project "TINY". We would also like to acknowledge Federico Lelli for insightful discussions. We also warmly thank Tobias Mistele for sharing the full data sets on the weak lensing rotation curves and for useful discussions.

\section*{Data Availability}

All the numerical scripts adopted in this work are available upon reasonable request to the corresponding Authors. The photometric and rotation curve data underlying this article are available from the \texttt{SPARC} database.\footnote{\url{https://astroweb.cwru.edu/SPARC/}}



\bibliographystyle{mnras}
\bibliography{example} 


\bsp	
\label{lastpage}
\end{document}